\documentclass[useAMS,usenatbib]{mn2e}
\usepackage{aas_macros}
\usepackage{graphicx}
\usepackage{hyperref}

\newcommand{\uvot}{\textit{Swift}/UVOT}

\newcommand{\xrt}{\textit{Swift}/XRT}

\newcommand{\po}{power-law}

\newcommand{\obj}{OJ\,287}

\title[Characterizing long-term optical, ultraviolet and X-ray variability in OJ\,287]{Characterizing long-term optical, ultraviolet and X-ray variability in different activity states of OJ\,287}

\author[ H. Siejkowski, A. Wierzcholska ]{
     H. Siejkowski$^{1}$\thanks{E-mail: h.siejkowski@cyfronet.pl}, A. Wierzcholska$^{2}$\thanks{E-mail: alicja.wierzcholska@ifj.edu.pl}, \\
     $^{1}$AGH University of Science and Technology, ACC Cyfronet AGH, ul. Nawojki 11, PO Box 386, PL-30-950, Krak\'{o}w 23, Poland\\
     $^{2}$Institute of Nuclear Physics, Polish Academy of Sciences, ul. Radzikowskiego 152, PL-31-342 Krak\'{o}w, Poland\\ 
}

\begin{document}

\date{Accepted .... Received ...; in original form ...}

\pagerange{\pageref{firstpage}--\pageref{lastpage}} \pubyear{2015}

\maketitle

\label{firstpage}

\begin{abstract}
    We have studied long-term optical, ultraviolet (UV)  and X-ray observations of \obj\ collected with the UVOT and XRT instruments mounted on board the {\it Swift} satellite to quantify spectral and temporal variability patterns observed during different activity states.
We characterized the flux variations using the data collected during almost 11 yr of the monitoring of the blazar.
Significant variability of the blazar has been detected both in the flux and spectral index   from the optical to X-ray regimes. We noted that the variability patterns observed in the optical range are more pronounced than the ones in the X-ray band. There is no clear relation between the optical/UV and X-ray emission, neither during the quiescence state nor during outbursts.
The most significant flares in the optical/UV regime were detected in 2015 December--2016 January. The shortest variability time-scale is one day and it is limited by the observation pointing.
A low activity state of \obj\  was observed at the end of 2014, while
 the beginning of 2015 revealed a flat X-ray spectrum, which has been observed for the first time. 
 On one hand, this can be a spectral upturn where the synchrotron and inverse Compton components meet, but on the other hand, it can be generated by an additional emission component.
The spectral studies  have not revealed any bluer-when-brighter or redder-when-brighter chromatism in the colour--magnitude diagram for \obj\ in any state of the source's activity.
A harder-when-brighter behaviour was noticed for \obj\ only in the case of the X-ray observations.
\end{abstract}


\section{Introduction}

Blazars, including BL Lacertae (BL Lac) type objects and flat spectrum radio quasars (FSRQs), represent a violent class of active galactic nuclei (AGNs). These are sources whose jets are pointing at a small angle with respect to the observer's line of sight \citep[e.g.][]{begelman84}.
Blazars are known to be highly variable  with variability observed at different time-scales from minutes  to years \citep[e.g.][]{wagner,2155flare,Gopal-Krishna11, saito,  Wierzcholska_s5}.
The observed emission extends from radio frequencies up to a high and very high energy regime \citep[e.g.][]{Wagner2009,  Abramowski2014, Wierzcholska_48}. The spectral energy distribution (SED) of blazars, in $\nu$--$\nu F_{\nu}$ representation, exhibits a double-humped structure.
The first, low-energy bump is usually attributed to the synchrotron emission of relativistic electrons from the jet, while the second is still a matter of debate, and leptonic, hadronic and hybrid scenarios are applied \citep[see e.g.][]{Maraschi92, Sikora94, Kirk98, Mucke13, Bottcher13}.

The two classes of blazars: BL Lac objects and FSRQs can be distinguished by different features in optical/UV spectra
\citep{Urry95}.
In the case of BL Lac type sources, featureless continuum emission is observed, while spectra of FSRQs are characterized with broad and narrow emission lines.
BL Lac type objects can be further  subdivided into high-,  intermediate- and low-energy-peaked BL Lac type objects (HBL, IBL, LBL, respectively) and this classification is based upon the position of the low-energy peak in the SED \citep[see, e.g.,][]{padovani95,fossati98,Abdo2010}.  Following \cite{Abdo2010}, for LBL sources the low-energy peak is located in the regime defined as $\nu_{\mathrm{s}}\leq10^{14}$\,Hz (infrared range), for IBL ones this regime is $10^{14}<\nu_{\mathrm{s}}\leq10^{15}$\,Hz (optical/UV range), while in the case of HBL blazars $\nu_{\mathrm{s}}>10^{15}$\,Hz (X-ray range). 
HBLs--IBLs--LBLs--FSRQs constitute a blazar sequence. This connection of decreasing bolometric luminosities and $\gamma$-ray dominance has been proposed and discussed  by \cite{fossati98} and  \cite{Ghisellini1998}
and has been updated into blazars envelope by \cite{Meyer11}.
Furthermore, \cite{Ghisellini2011} studied the properties of SEDs and emission lines of a large sample of sources from a one-year all-sky survey by the
\textit{Fermi} satellite and  proposed a physical distinction between FSRQs and BL Lac objects  based on the luminosity of the broad line region.

\obj\ ($z=0.306$) has become one of the best monitored blazars in the optical regime since its identification in this range by \cite{Dickel67}. The source is classified as an LBL type object. Deeper studies of the source started in the 1980s and therefore \obj\ is a perfect candidate for long-term variability studies and searching for periodicities.
An almost 12-yr periodicity is observed in the optical monitoring of \obj\ with two peaks observed during every flaring event \citep[e.g.][]{Valtonen06}.
Additionally, a 60-yr variability was claimed by \cite{Valtonen06}.
This behaviour is usually explained in terms of the binary black hole model \citep[see e.g.][]{Sillanpaa88, Sillanpaa96}.
An indication for a shorter periodic behaviour in the optical range was suggested by \cite{Sagar04}, \cite{Wu06}, \cite{Gupta12}, and \cite{Valtonen12}.

Despite many years of observations, \obj\  has not been well studied in the other energy regimes except for the optical and radio ones. 
This work presents 12 yr of multifrequency observations of \obj\ preformed with the XRT and UVOT instruments onboard the {\it Swift} space telescope. The paper is organized as follows:
Section~\ref{data} presents the data analysis details. Section~\ref{temporal} and \ref{spectral} focus on a characterization of the temporal and spectral variability of \obj, respectively.
The work is summarized in Section~\ref{summary}.

\section{Data analysis} \label{data}

\subsection{X-ray observations}
X-ray observations made with the XRT telescope on board the {\it Swift} satellite  (see \citealt{Gehrels04} for details) in the energy range of 0.3--10\,keV were analysed using version  6.19 of the \textsc{heasoft} package. In these studies, all observations made with \xrt\ in the PC mode from the mission start up to MJD 57552.93053 are considered. The total number of observations analysed is 280. 
All the data were reprocessed using the standard \textsc{xrtpipeline} procedure\footnote{\url{www.swift.ac.uk/analysis/xrt/}}. In the case of each observation for the spectral fitting, \textsc{xspec} (v.12.9.0n) was used \citep{Arnaud96}.  All light-curve points were derived by fitting the power-law model with the value of the Galactic absorption frozen at $N_\mathrm{H} = 2.56 \times 10^{20}\,\textrm{cm}^{-2}$ taken from \cite{Kalberla05}. The observations were corrected for the pile-up in the PC mode whenever the count rate  was 0.5 or higher.

\subsection{Optical and ultraviolet observations}
Simultaneously with the XRT instrument, \obj\ was monitored with \uvot\ in the optical/UV band. 
The optical and UV observations were taken in six bands: \textit{UVW2} (188\,nm), \textit{UVM2} (217\,nm),  \textit{UVW1} (251\,nm),  \textit{U} (345\,nm), \textit{B} (439\,nm) and \textit{V} (544\,nm). The \verb|uvotsource| procedure was used in order to calculate the instrumental magnitudes in the aperture with a radius of 5\,arcsec. 
The background area was defined as a circular region with a radius of 5\,arcsec located close to the source region and not being  contaminated with any signal from the
nearby sources. 
The influence of three different sizes of background was checked and the results obtained were consistent within the uncertainties.
All magnitudes were converted into fluxes using the conversion factors provided by \cite{Poole08}.
The data were corrected for dust contamination using the reddening $E(B-V)=0.0241$ from \cite{Schlafly11} and ratios of the extinction to reddening ratios from \cite{Giommi06}.

\section{Temporal variability studies} \label{temporal}
\begin{figure*}
\centering{\includegraphics[width=0.98\textwidth]{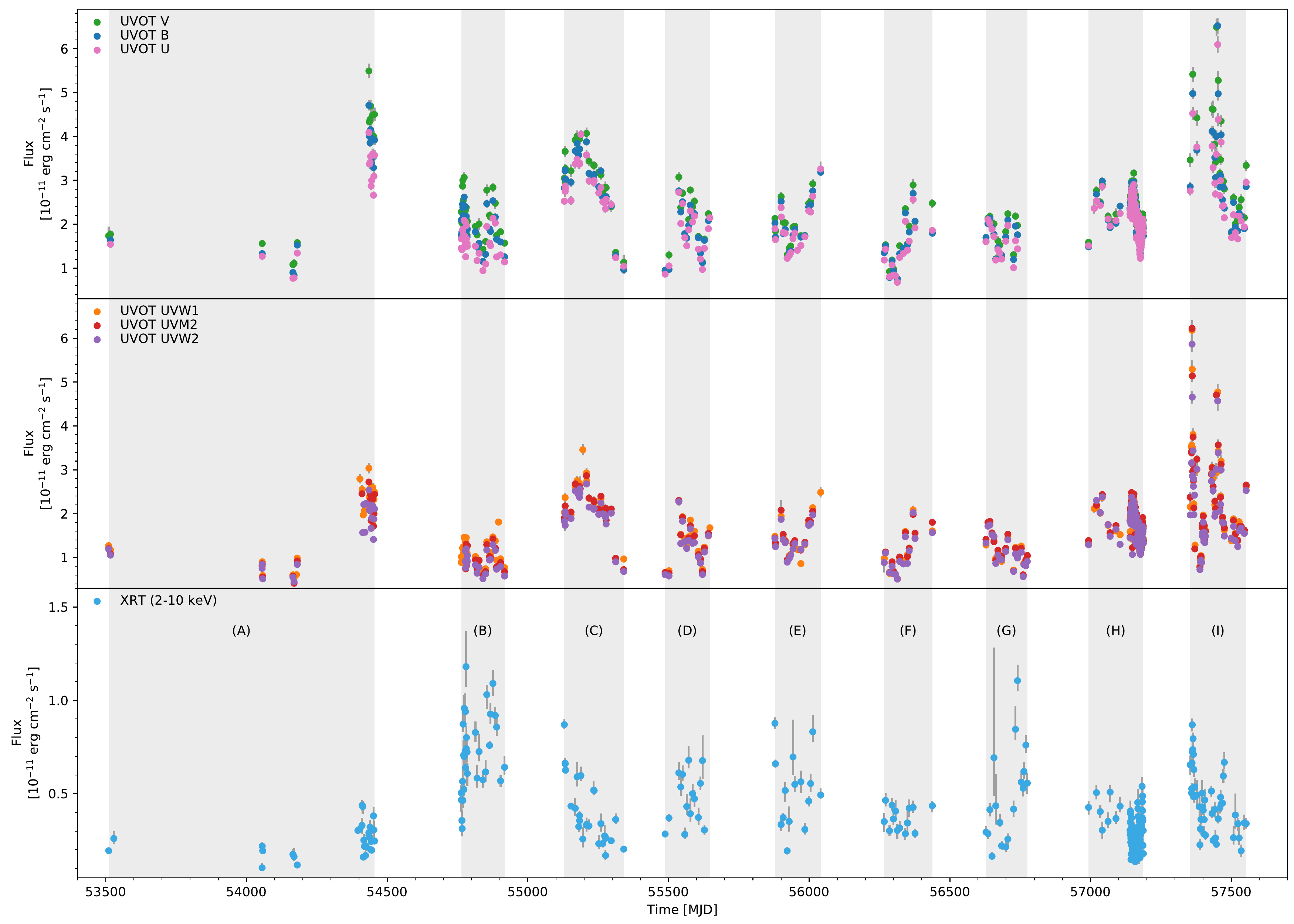}}\\
\caption{The long-term light curve presenting the optical, UV and X-ray observations of \obj\ made during the period of 2005--2016. The vertical grey areas indicate the A--I intervals chosen for detailed studies. }
\label{lc_uvot_xrt}
\end{figure*}

During the period of 2005--2016, \obj\ was observed many times with \uvot\ and \xrt\, resulting in 280 pointing observations. The long-term optical, UV and X-ray light curve of the source is presented in Fig.~\ref{lc_uvot_xrt}. The upper panel of the plot shows the optical observations performed with \uvot\ in the \textit{U}, \textit{B}, and \textit{V} filters; the middle one shows the UV observations made in \textit{UVW1}, \textit{UVM2} and \textit{UVW2} filters; and the bottom one shows the X-ray monitoring performed with \xrt\ in the energy range of 2--10\,keV. 
The energy range chosen allowed us to compare the results obtained in previous works focusing on an X-ray data analysis.
For the temporal variability studies, we used the entire set of data as well as nine intervals marked in  Fig.~\ref{lc_uvot_xrt} with vertical grey areas and  letters A--I. Each interval represents a similar flux level and consists of series of consecutive observations.

\begin{figure}
\centering{\includegraphics[width=0.49\textwidth]{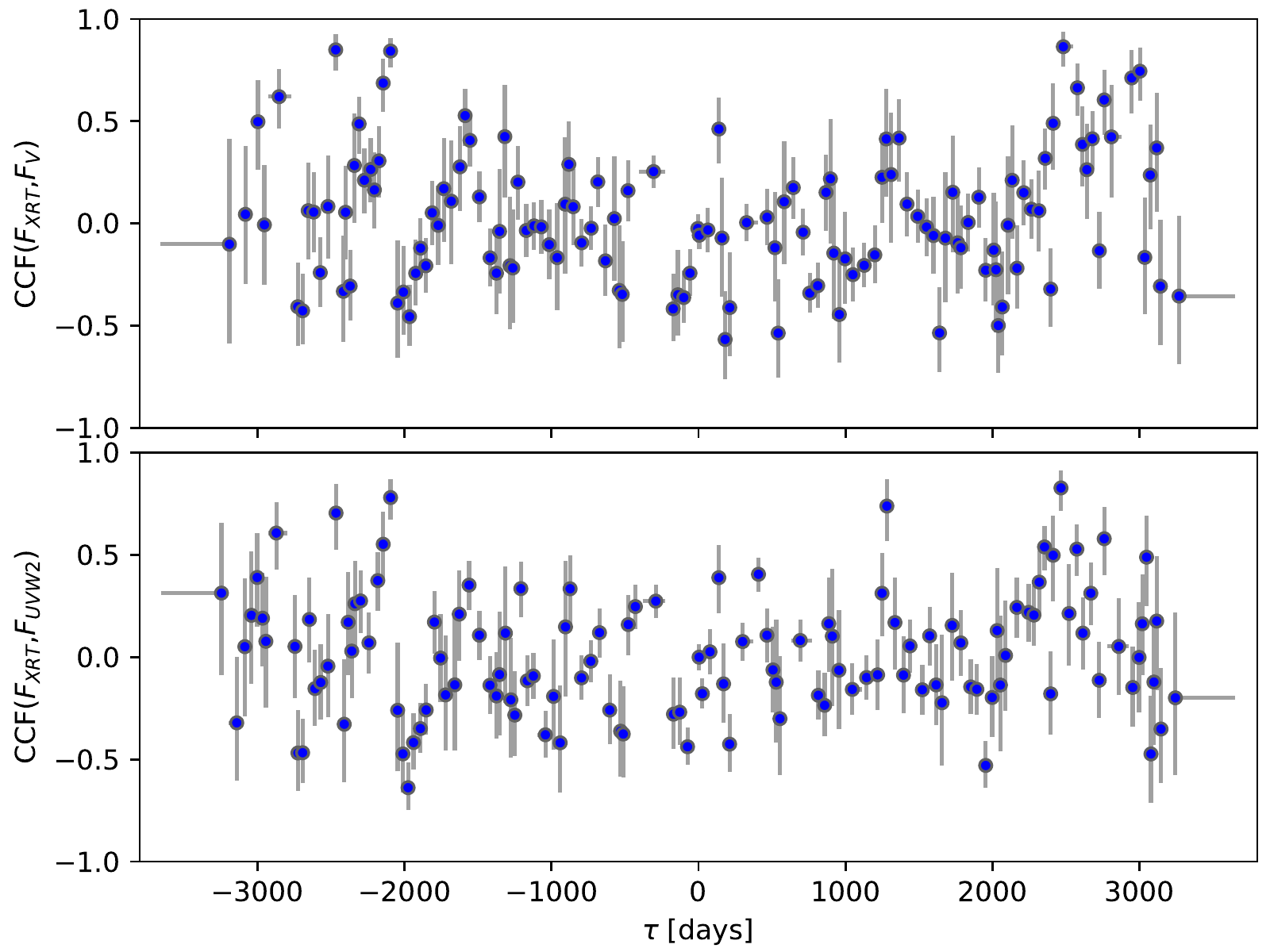}}\\
\caption{The DCF between the X-ray flux ($F_\mathrm{XRT}$) and the filter \textit{V}flux ($F_V$) and filter \textit{UVW2} flux ($F_{UVW2}$).}
\label{dcf}
\end{figure}

Strong flux changes were observed in the X-ray data as well as in the optical and UV range. 
A comparison of the flux light curves for the data collected in the X-ray as well as in the optical and UV regimes does not show any clear relation between the different energy ranges. Strong optical/UV outbursts did not have so strong counterparts in the X-ray regime and inversely. The discrete correlation function \citep[DCF,][]{Edelson1988} does not reveal any relation between the optical/UV and X-ray observations (Fig.~\ref{dcf}). In order to obtain the DCF, we used the algorithm implemented by \cite{Alexander97}.

Let us notice here that \obj\ is a LBL type blazar and in the case of such a source the optical/UV range corresponds to the first, low-energy  bump in the SED, while the high energy one is placed in the X-ray regime. This causes that the changes observed in the optical/UV regime and the X-ray one may be caused by different physical processes responsible for the emission observed. 

\begin{figure}
\centering{\includegraphics[width=0.48\textwidth]{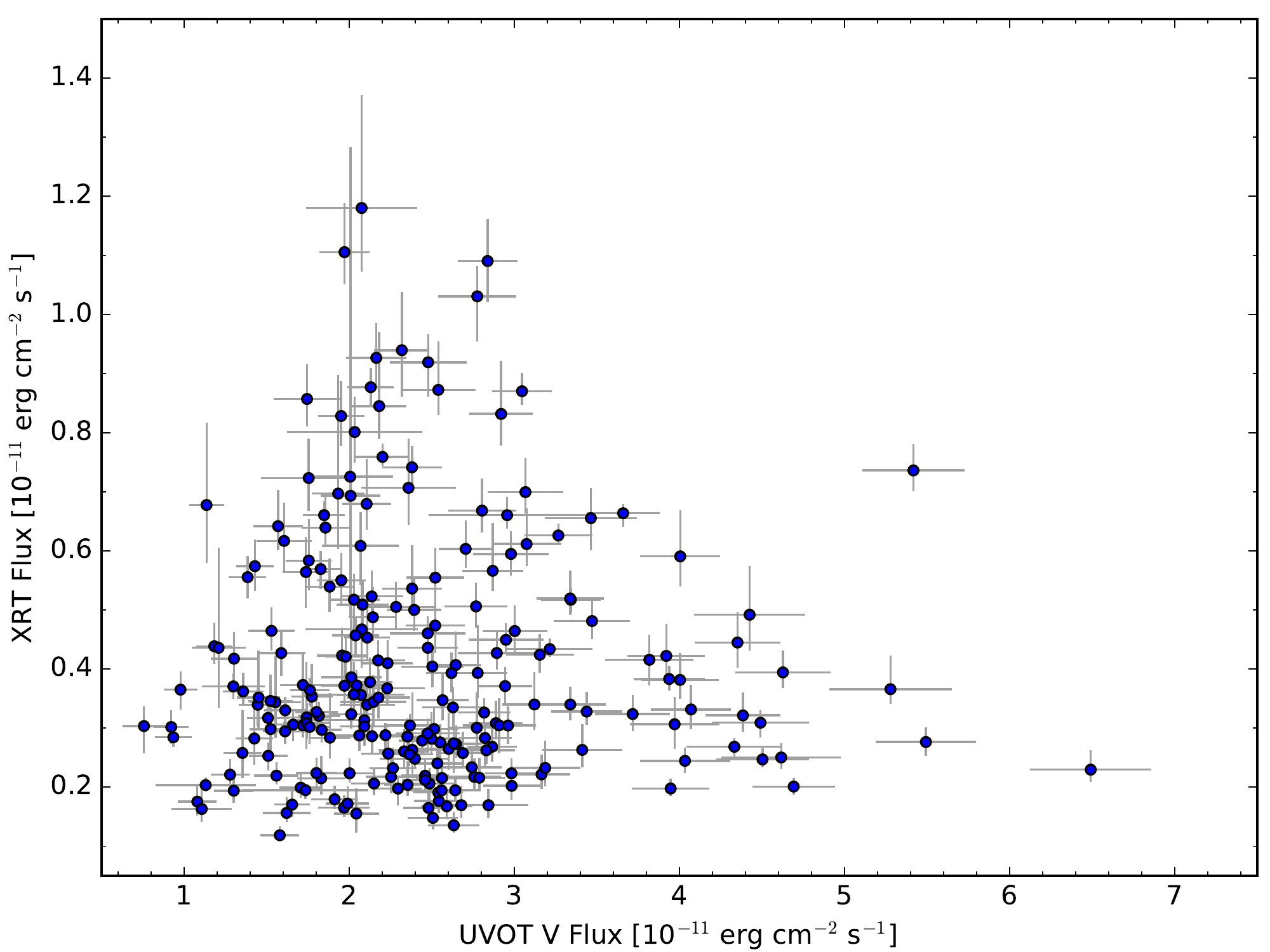}}\\
\caption{A comparison of the X-ray and optical fluxes collected in the \textit{V} band. }
\label{xrt-vs-uvot-V}
\end{figure}

Fig.~\ref{xrt-vs-uvot-V} shows a comparison of the X-ray and optical \textit{V} band fluxes. The result presented in the plot confirms no  relation visible in the light-curve plot (see Fig.~\ref{lc_uvot_xrt}). Pearson's correlation coefficient for this case is $C=-0.005$, which indicates
no clear correlation between these two ranges. The same result, a lack of any clear relation, has been obtained for the comparison of the X-ray band with every optical/UV band.

As a first step of a characterization of the UV, optical and X-ray long-term variability of the blazar, the  fractional variability amplitude was used. It is defined by \cite{Vaughan2003} as follows:
\begin{equation}
 F_\mathrm{var}= \sqrt{\frac{S^2-err^2}{F^2}},
\end{equation}
where $F$ is the mean flux and $S^2$ is its variance and $err^2$ is the mean square error.
The errors of the fractional variability amplitude are calculated following the formula by \cite{Poutanen08}:
\begin{equation}
  \delta  F_\mathrm{var}= \sqrt{F_\mathrm{var}^2+(\sigma^2)} -F_\mathrm{var},
\end{equation}
where the error in the normalized excess variance $\sigma$ is calculated following \cite{Vaughan03}:
\begin{equation}
 \sigma = \sqrt{\left( \sqrt{\frac{2}{N} }\frac{err^2}{F^2} \right)^2   + \left( \sqrt{\frac{err^2}{N}}\frac{2 F_\mathrm{var}}{F} \right)^2        },
\end{equation}
where $N$ indicates the number of data points in the light curve. 

In the next step, the doubling/halving time-scale was used in order to characterize the variability of \obj. The quantity is defined in two ways \citep[following e.g.][]{Zhang99}: 
\begin{itemize}
    \item[(i)] as the smallest value of the set constituted of $(k,m)$ pairs:
\begin{equation}
    \tau_{k,m}=\left|\frac{\Phi \, \Delta T}{\Delta \Phi}\right|=\left|\frac{(\Phi_k + \Phi_m ) \, (T_k-T_m)}{2 \, (\Phi_k-\Phi_m)}\right| \mathrm{and}
\end{equation}
 \item[(ii)] the mean value of the five smallest items from $\{\tau_{k,m}\}$.
\end{itemize}
The flux value at the time of $T_j$ is denoted by $\Phi_j$, for all pairs of points $(k,m)$ in the light curve. The doubling/halving time-scale for the definition (i) is marked as $\tau^{(i)}_\mathrm{d}$, while for (ii) as $\tau^{(ii)}_\mathrm{d}$. The latter value can be used as a cross check for the $\tau^{(i)}_\mathrm{d}$. Single minimum value can be generated by chance, and the $\tau^{(ii)}_\mathrm{d}$ gives information how small is the minimum compared to the five smallest values (mainly due to irregular sampling). If the $\tau^{(i)}_\mathrm{d}$ is much different from the $\tau^{(ii)}_d$ then the probability that the minimum is accidental is high.

The variability parameters are collected in Table~\ref{variability_xrt} and \ref{variability_uvot} for the X-ray and optical/UV observations, respectively.
In addition to the fractional variability amplitude and doubling time-scale, the tables present the mean flux value and the reduced $\chi^2$ of the fit with a constant value.
All quantities were calculated for all the data collected during the period between 2005 and 2016 and for the nine separate intervals (A--I).

The fractional variability amplitude and  $\chi^2$ value confirm significant variability observed in all the energy ranges studied.
The most prominent variability for all optical and UV filters is observed for interval~A. The fractional variability amplitude for this interval is between 0.44 and 0.49. In the case of the X-ray monitoring, the largest value of $F_\mathrm{var}$ is for the entire set of data.
The doubling time-scale values calculated determine the shortest variability time-scale observed in the long-term monitoring of \obj\ as one day.

\section{Spectral variability studies} \label{spectral}

\begin{figure}
\centering{\includegraphics[width=0.48\textwidth]{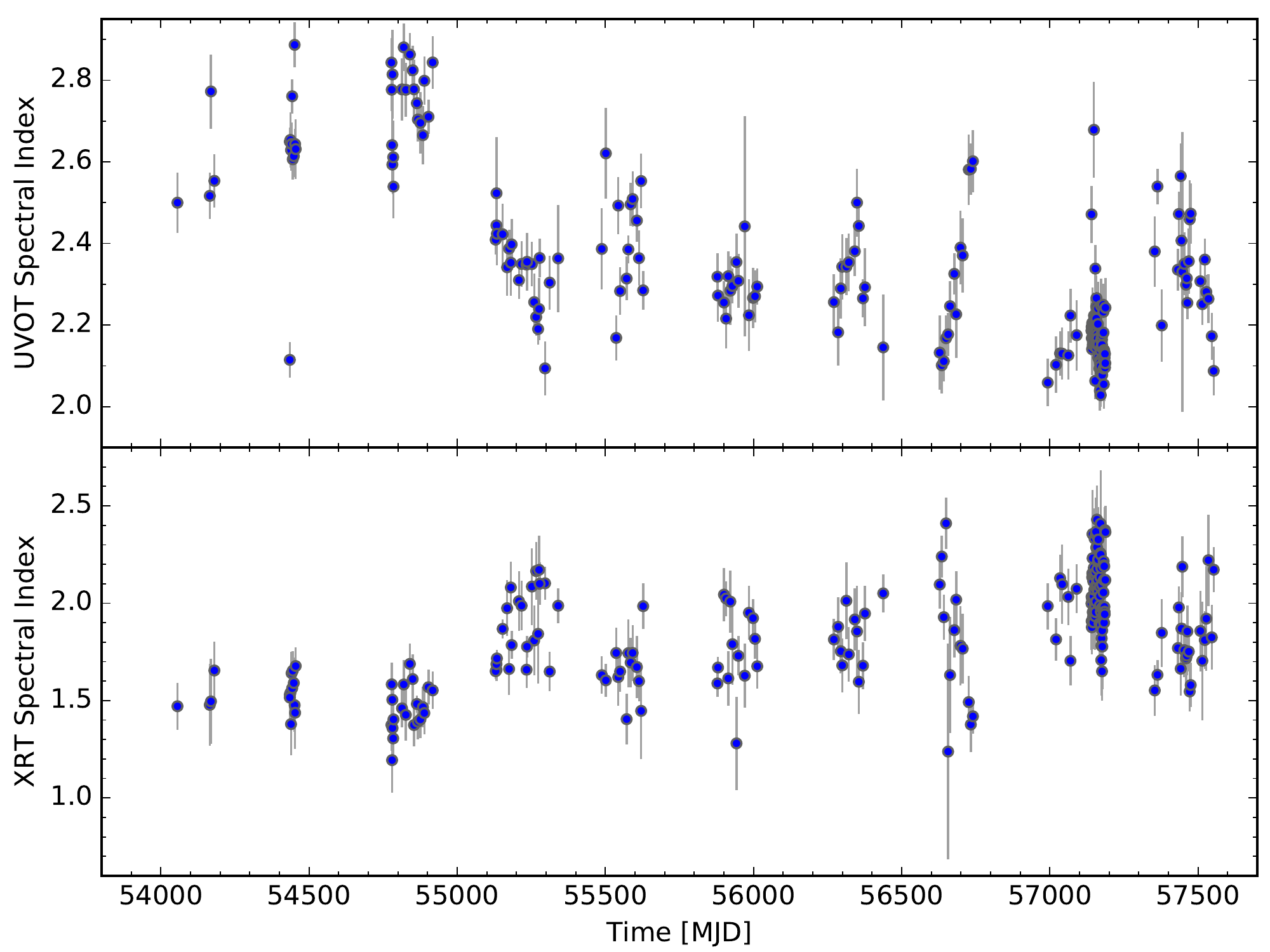}}\\
\caption{The time evolution of the long-term spectral indices for the optical/UV range (top panel) and for the X-ray band (bottom panel). The optical/UV spectral index is found by fitting a power law to data points (for details see Sect.~\ref{spectral}).}
\label{lc_indices}
\end{figure}

\begin{figure}
\centering{\includegraphics[width=0.48\textwidth]{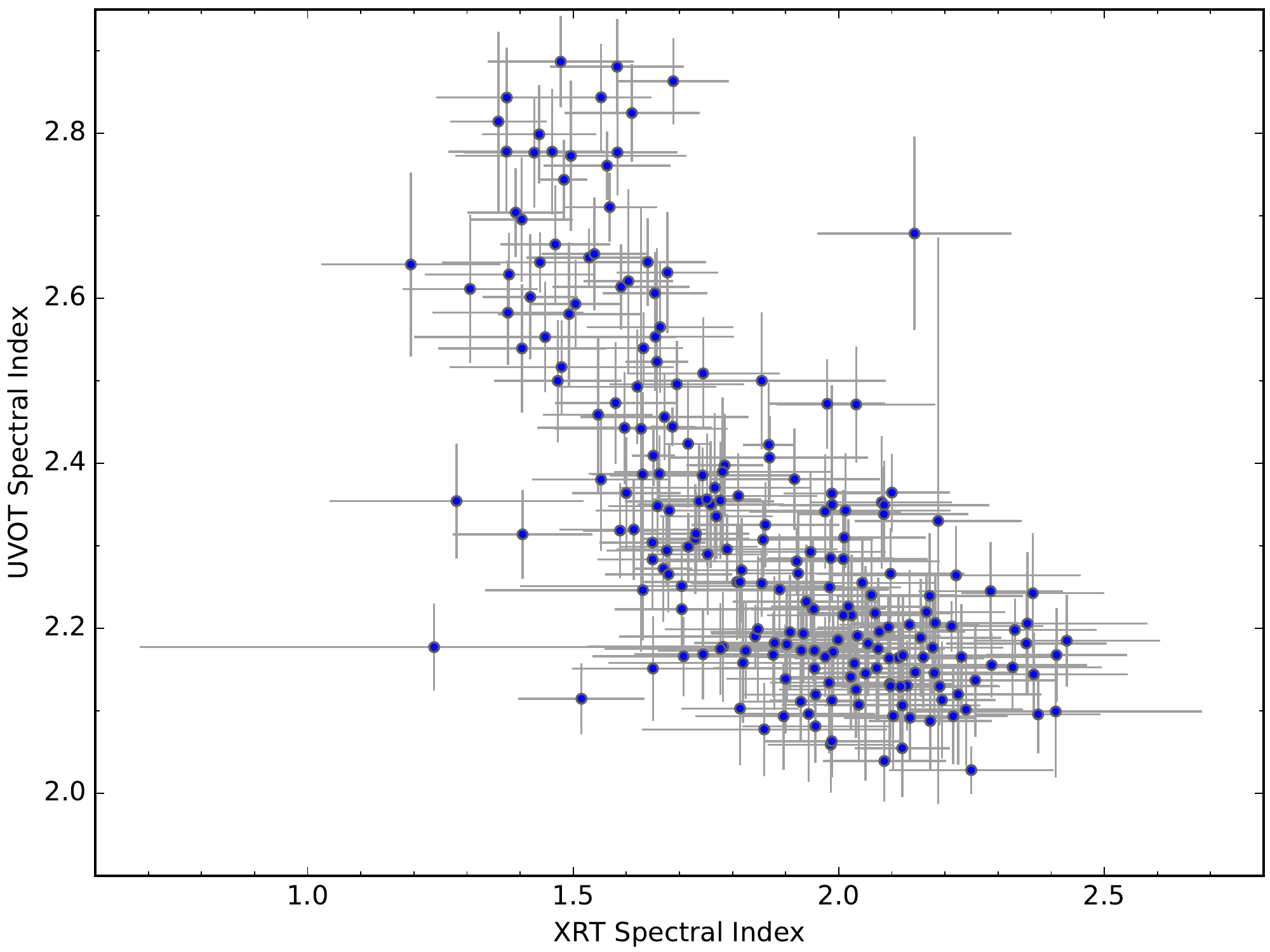}}\\
\caption{A comparison of the spectral indices obtained from X-ray and optical/UV observations.}
\label{indices}
\end{figure}

For the optical/UV observations in order to  derive the spectral index of emission in this band, we fitted a power law to the data gathered with \uvot\ for each ObsID. 
We only used observations that have measurements in each filter for a given ObsID (i.e. six data points). 
The fitting was done with the  \texttt{optimize.curve\_fit} 
function from the \textsc{scipy} version 0.18.0 package \citep{scipy}, which uses the Levenberg--Marquardt algorithm.
Fig.~\ref{lc_indices} shows the evolution in time of the X-ray and optical/UV spectral indices, while Fig.~\ref{indices} presents a plot of the X-ray versus optical/UV spectral index.
The spectral index in the case of the X-ray observations was obtained with a \po\ fit to a single observation. 

In the case of the X-ray observations, the spectral index varies between 1.19 and 2.43, while for the optical/UV ones, the range is 2.03--2.89.
The plots show significant variability in the spectral index's evolution for all wavebands. The changes observed both in the spectral index's evolution and the correlation plots show a strong anti-correlation between these two quantities. The correlation coefficient for this comparison is $C=-0.71\pm0.04$.  The uncertainty of Pearson's correlation coefficient was estimated using a Monte Carlo approach following \cite{Wierzcholska_48}.

\begin{figure}
\centering{\includegraphics[width=0.48\textwidth]{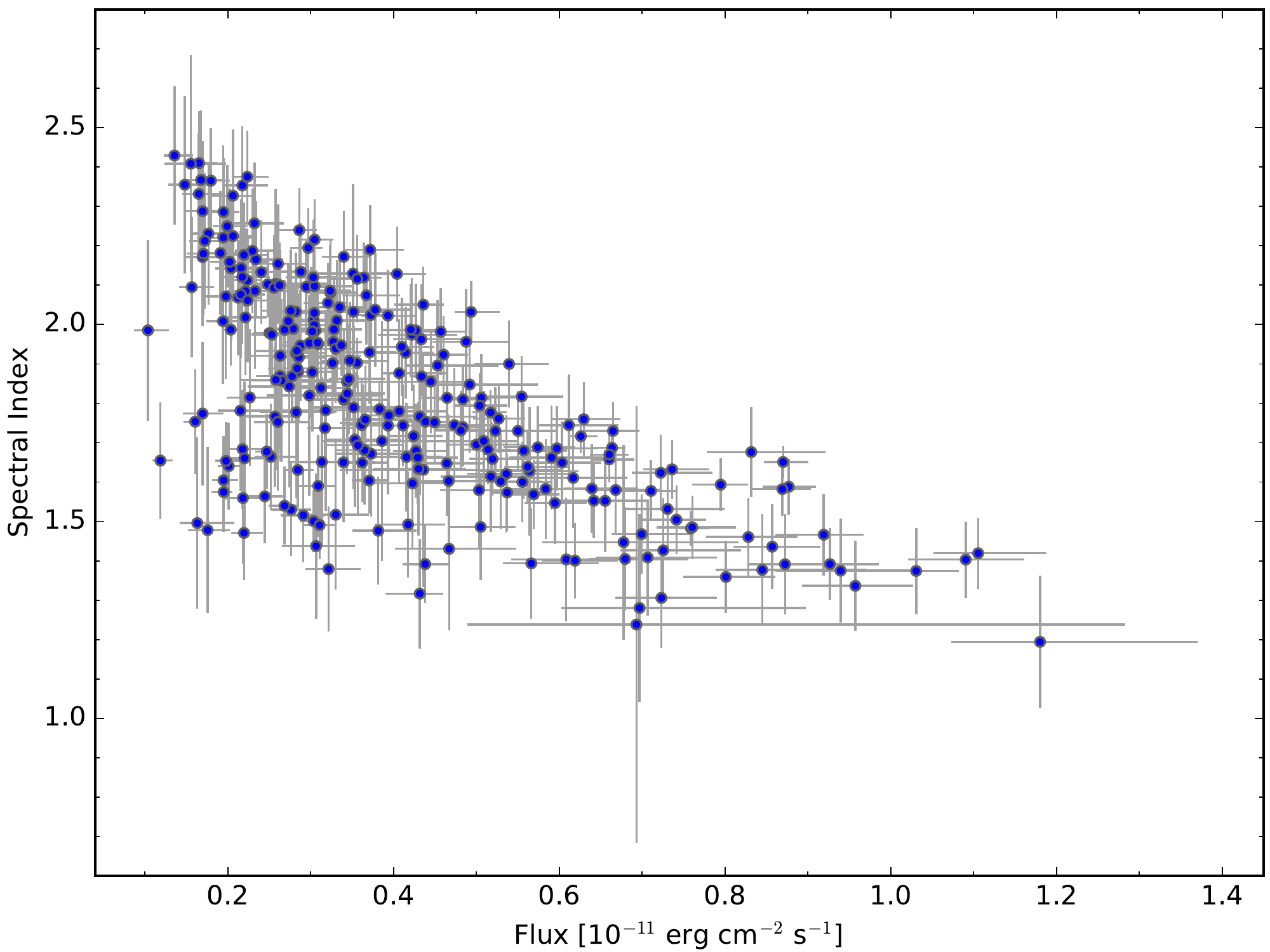}}\\
\caption{A comparison of the X-ray flux and the corresponding spectral index obtained from the \po\ fit.}
\label{xray-flux-vs-ind}
\end{figure}

We also compare the X-ray spectral index with the corresponding flux (see Fig.~\ref{xray-flux-vs-ind}). The correlation visible in the plot, characterized with Pearson's correlation coefficient of $C=-0.67\pm0.02$, indicates a harder-when-brighter behaviour in the long-term observations. This  spectral hardening with the increasing flux is a typical feature of HBL type sources \citep[see e.g.][]{Pian88, Zhang2005}.

\begin{figure}
\centering{\includegraphics[width=0.48\textwidth]{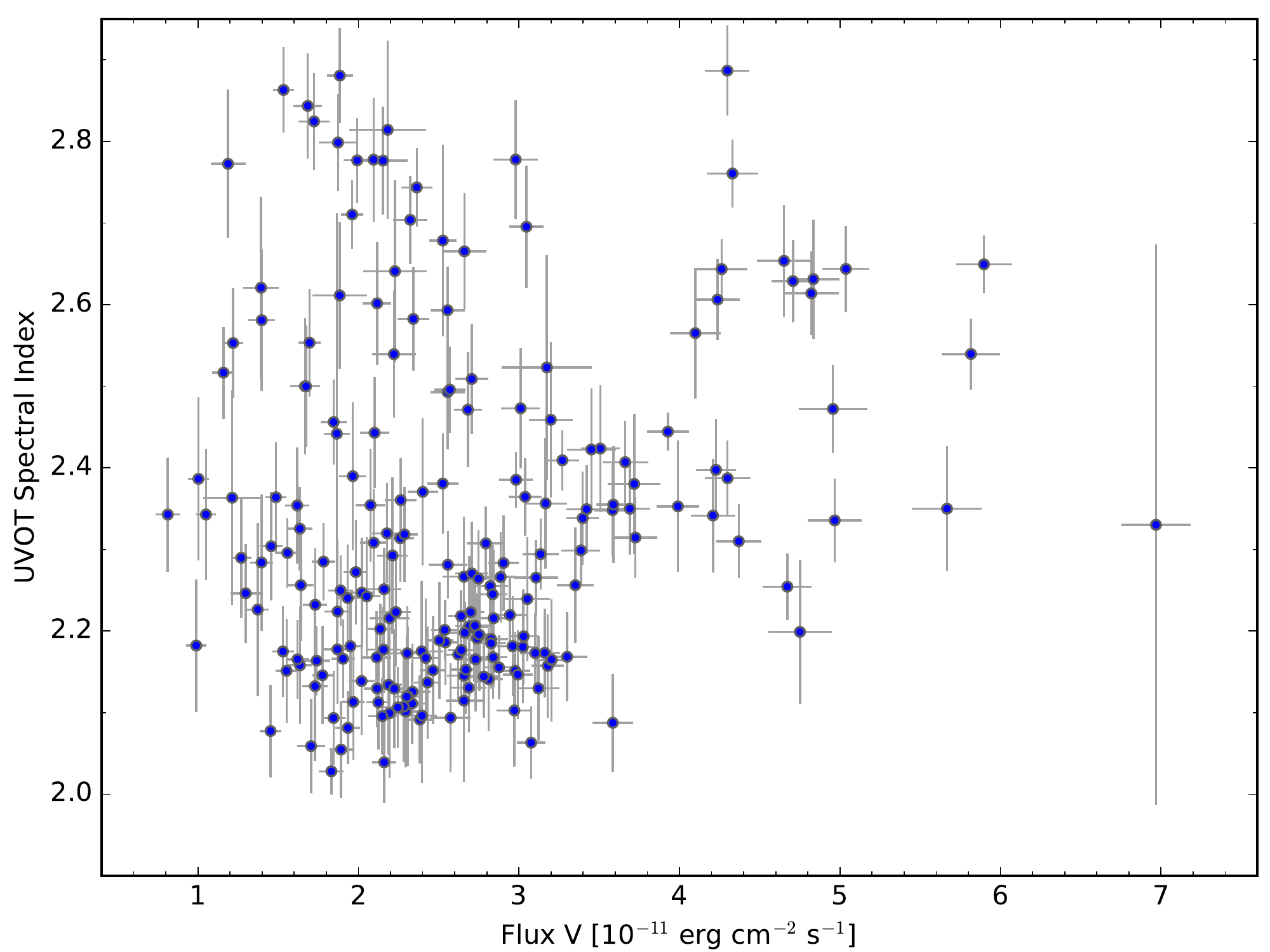}}\\
\centering{\includegraphics[width=0.48\textwidth]{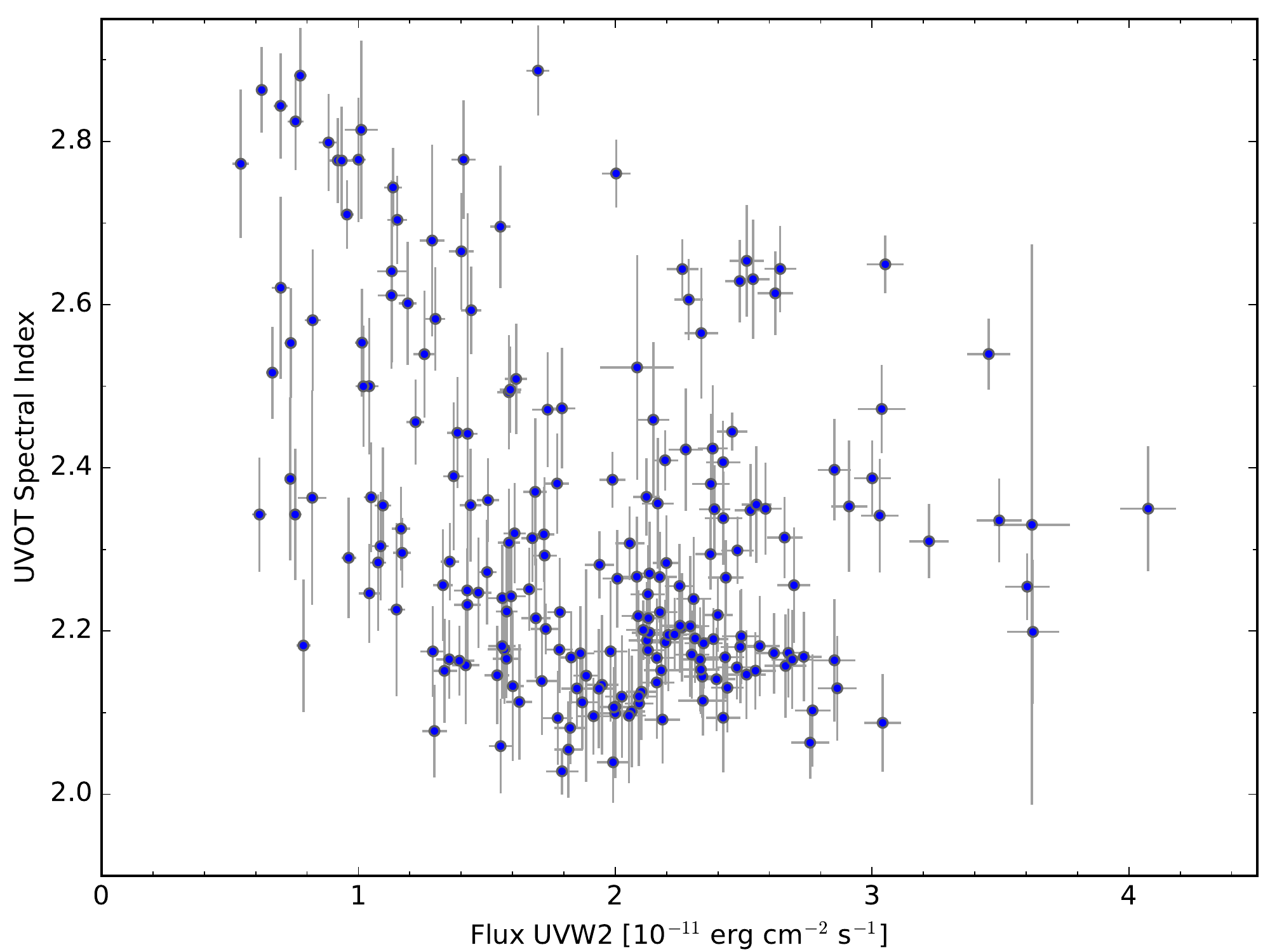}}\\
\caption{A comparison of the optical flux in the \textit{V} filter (top panel) and the optical/UV spectral index obtained from the \po\ fit (see Sect.~\ref{spectral}). The bottom panel shows an analogical relation but for the \textit{UVW2} filter.}
\label{uvot_fl_vs_ind_V}
\end{figure}

We have also noted a lack of correlation between the fluxes observed in the X-ray and optical/UV regimes. This is in  agreement with the absence of a harder-when-brighter relation in the flux-index diagram for the optical observations (see Fig.~\ref{uvot_fl_vs_ind_V}).

\subsection{Colour--magnitude diagram}

\begin{figure}
\centering{\includegraphics[width=0.48\textwidth]{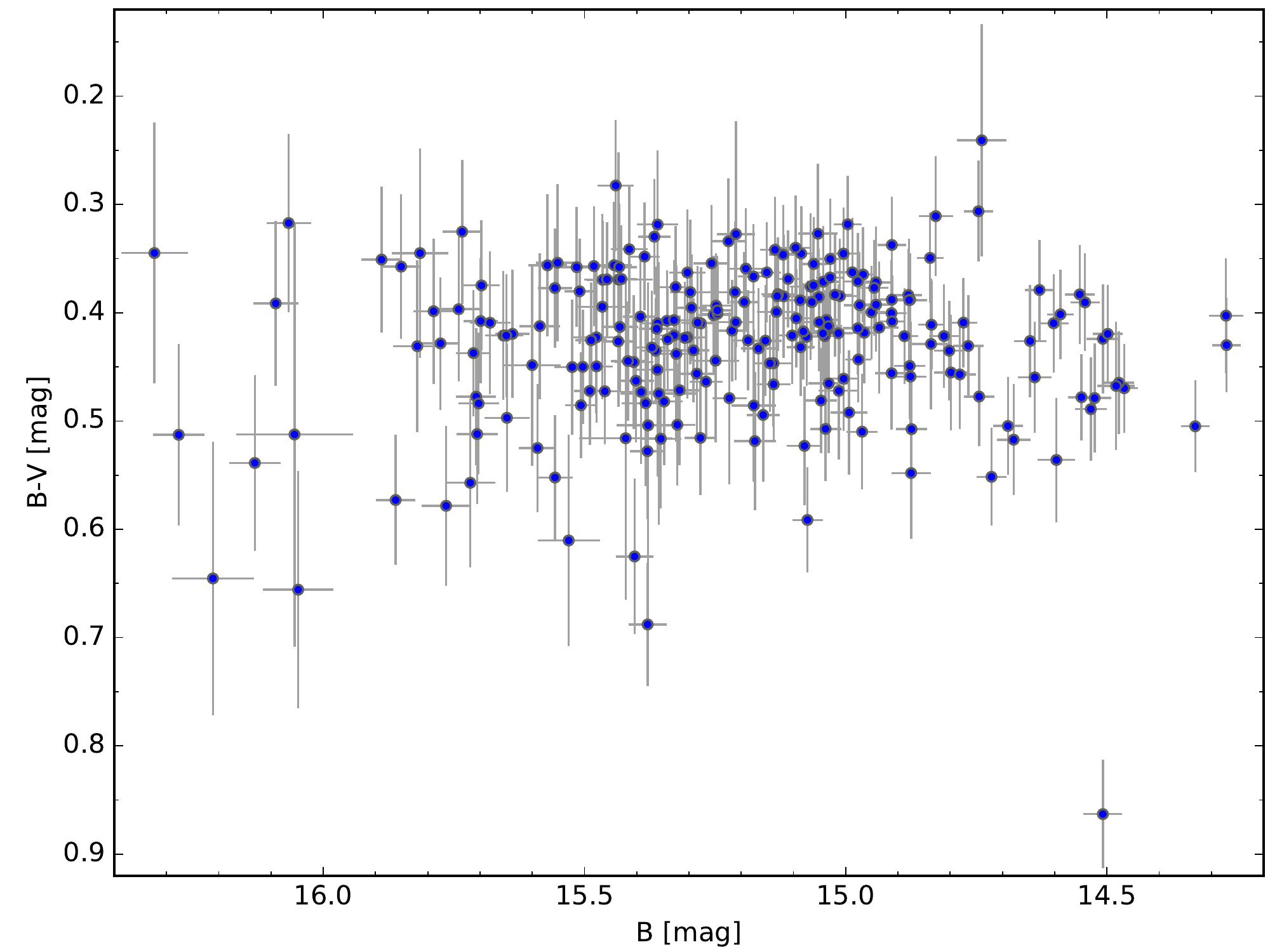}}\\
\centering{\includegraphics[width=0.48\textwidth]{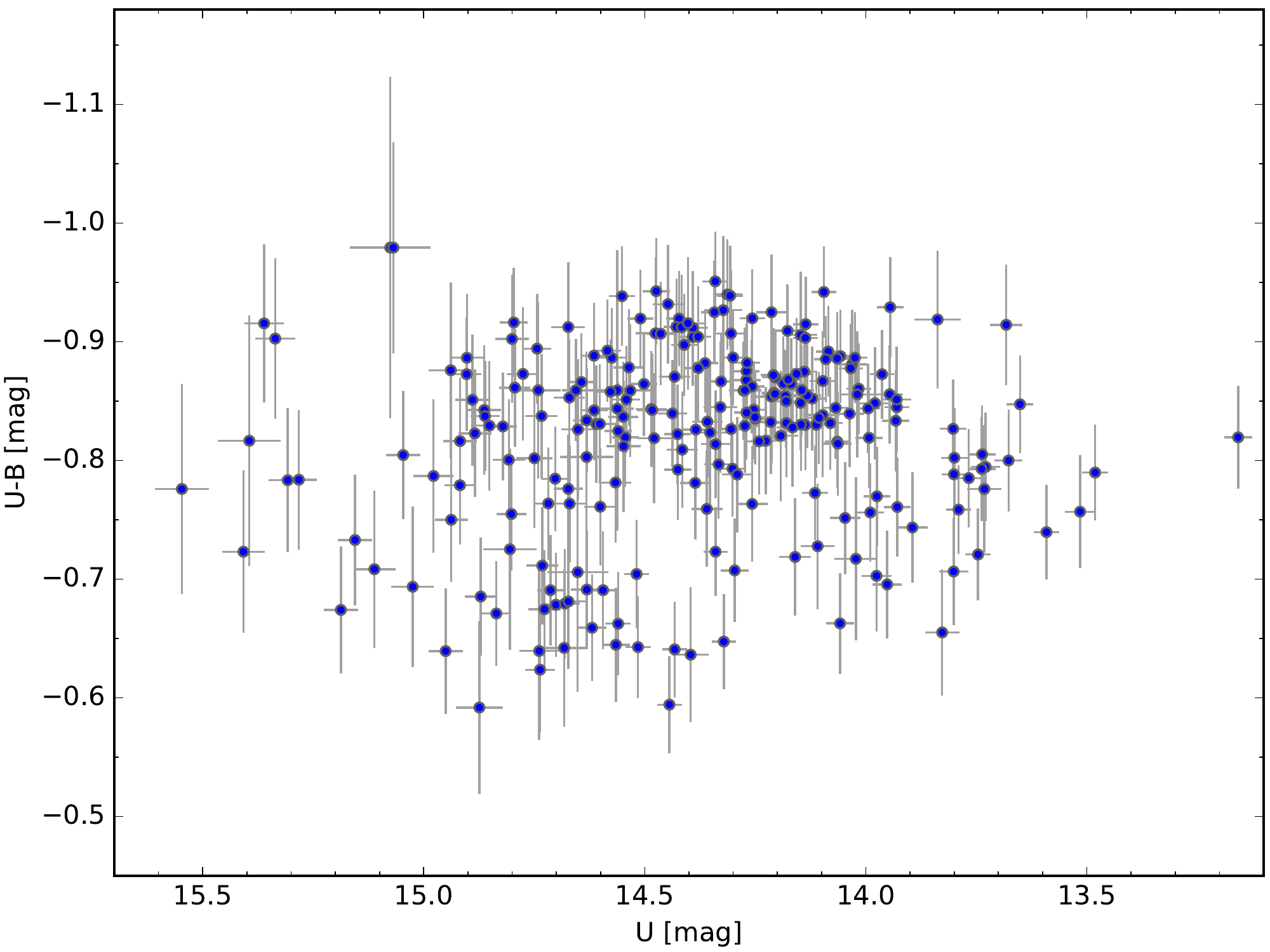}}\\
\centering{\includegraphics[width=0.48\textwidth]{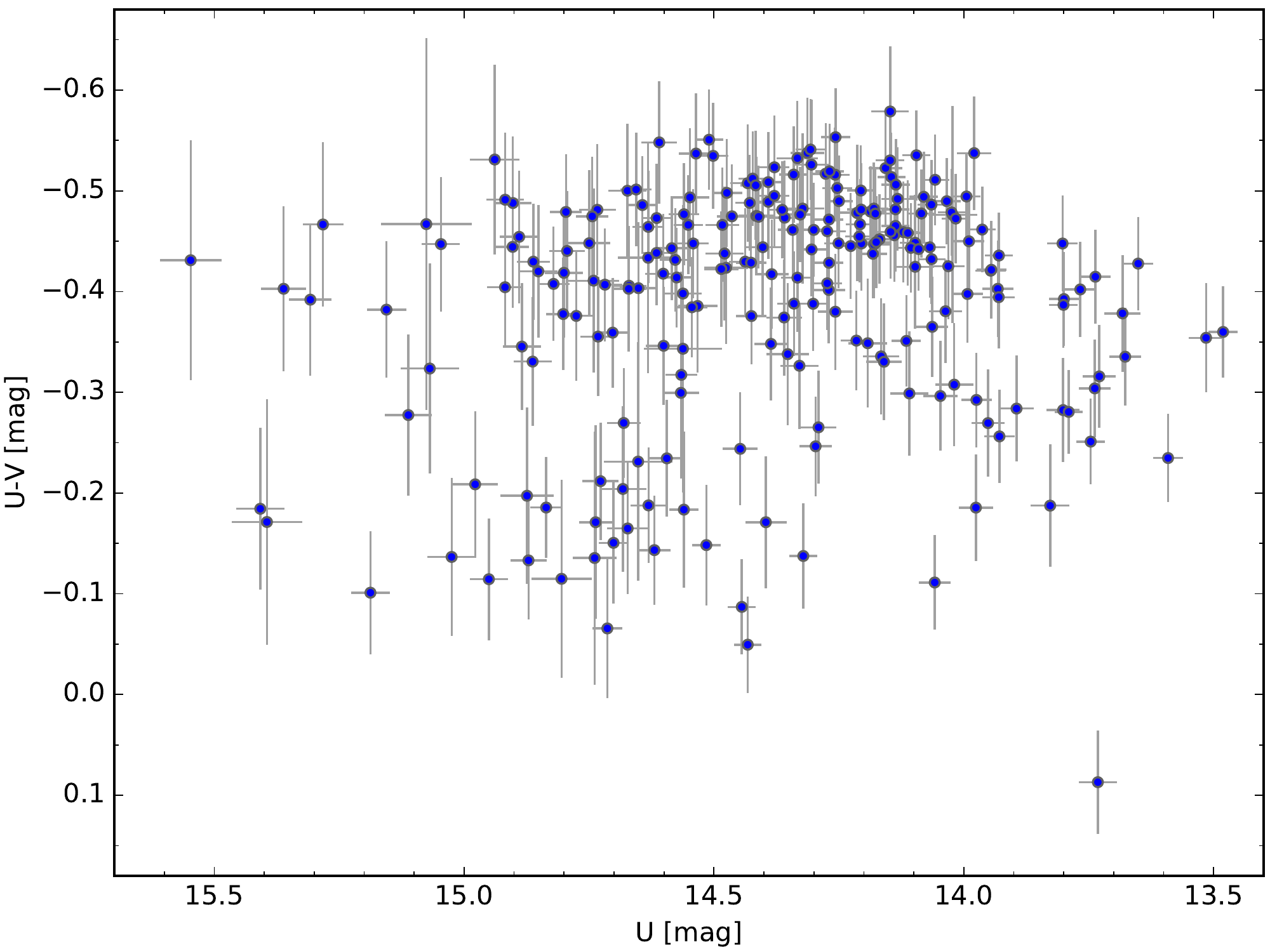}}\\
\caption{Colour-magnitude diagrams for the optical observations performed with \uvot. The plots present a comparison of: (\textit{B}$-$\textit{V}) versus \textit{B}, (\textit{U}$-$\textit{B}) versus \textit{U}, (\textit{U}$-$\textit{V}) versus \textit{U}. }
\label{col_mag}
\end{figure}

Three colour--magnitude diagrams, namely (\textit{B}$-$\textit{V}) versus \textit{B}, (\textit{B}$-$\textit{U}) versus \textit{B}, (\textit{V}$-$\textit{U}) versus \textit{V}, are presented in Fig.~\ref{col_mag}. 
For each case, the Pearson's correlation coefficients were found to be equal to 0.05, 0.11 and 0.13 for the corresponding colour--magnitude diagrams, respectively. 
The correlation coefficients found suggest that there is no significant bluer-when-brighter or redder-when-brighter trend in the entire set of the observations discussed.
Furthermore, the analysis of shorter intervals of observations both in the low and high states also did not show any chromatism observed in the \uvot\ observations.

Previous works focusing on the colour--magnitude relation in \obj\ have revealed various types of behaviours. Some hints for a bluer-when-brighter chromatism were reported by \cite{Carini1992} while studying observations collected in the \textit{V} and \textit{B} bands in 1973--1976.
A clear bluer-when-brighter chromatism during a flaring state was found by \cite{Dai2011}. The observations mentioned have been collected in the \textit{R} and \textit{V} bands from 1993 to 1997.
Similarly, \cite{Ikejiri_2011} found an indication for a weak colour--magnitude correlation in the observations collected from 2008 May to 2010 January in the \textit{V} and \textit{J} filters.
A lack of a bluer-when-brighter or redder-when-brighter relation was noticed for shorter time-scales, as well as for long-term observations collected in the \textit{B} and \textit{R} bands during the period 2007--2012 with the ATOM telescope \citep{Wierzcholska_atom}.

The various behaviour of the optical colour as a function of the magnitude observed  of \obj\ indicates a complex behaviour of the optical emission mechanisms at work in this blazar. An absence of a bluer-when-brighter chromatism can be caused by multiple episodes characterized with different bluer-when-brighter slopes.
In such a case, the relation cannot be visible in a large set of observations, but can be detected in shorter periods.

\subsection{Optical-UV-X-ray SEDs}
\begin{figure*}
\centering{\includegraphics[width=0.98\textwidth]{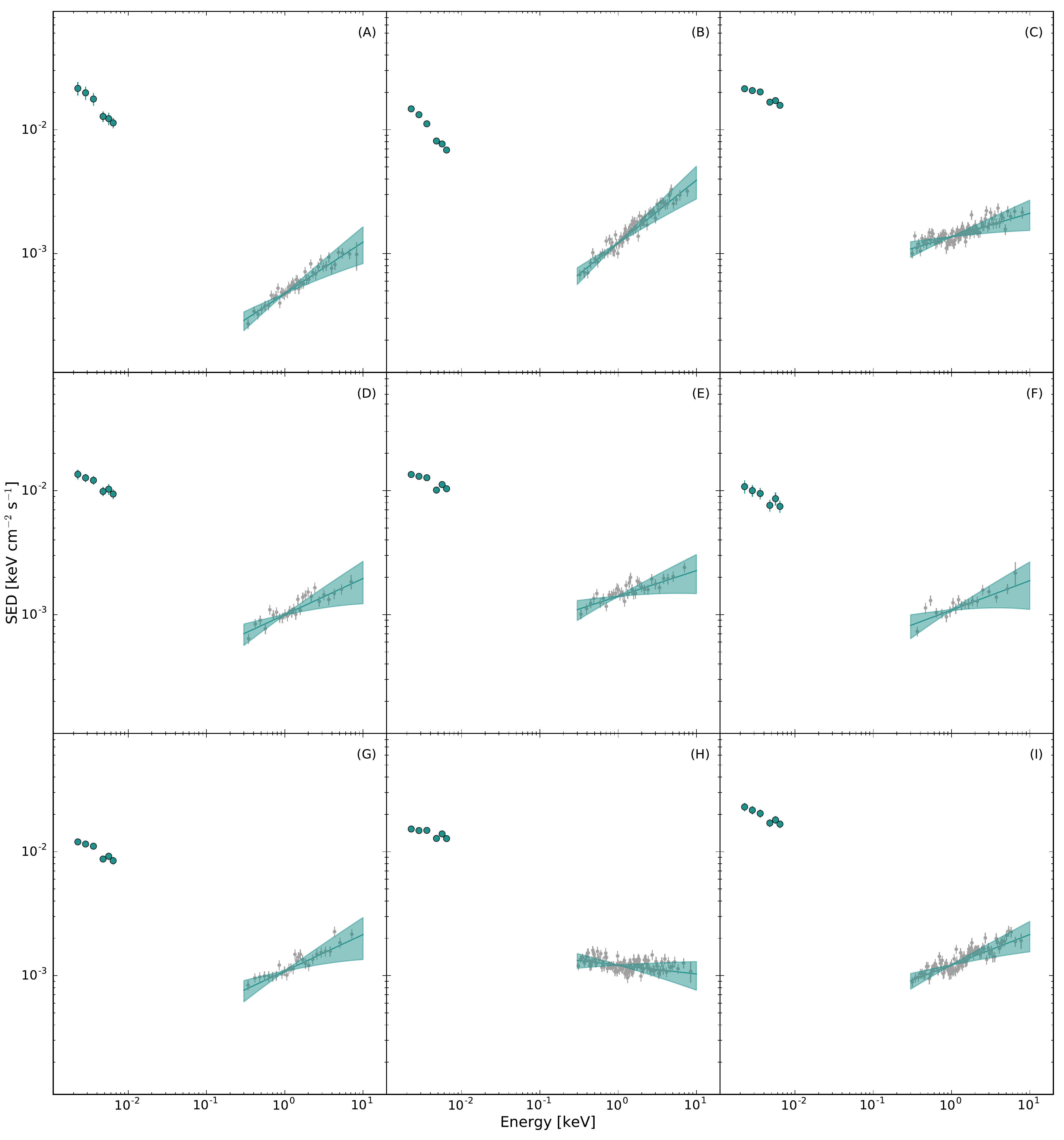}}\\
\caption{SED for \obj\ for the intervals defined in Sect.~\ref{temporal}. The colour points represent the optical/UV observations and the grey points -- the X-ray observations. The butterfly plots show the \po\ model with the uncertainty of the fit at a confidence level of 1$\sigma$.}
\label{intervals_seds}
\end{figure*}

Fig.~\ref{intervals_seds} shows the SED for the nine intervals defined in Sect.~\ref{temporal}. The parameters of the X-ray spectra are listed in Table~\ref{tab:xrt-spec-intervals}.
All of the nine cases show that the soft optical/UV spectrum corresponds to the hard X-ray one, and vice versa. Moreover, the following pattern emerges: The harder the optical/UV spectra, the softer the X-ray spectra.
We also note that in the case of the H interval, the X-ray spectrum is characterized with a spectral index of about 2.0. We consider two possible scenarios explaining such an evolution of the X-ray spectrum. On one hand, this spectral shape can be caused by the fact that the X-ray regime is a place where two spectral components meet and the flat spectrum is a consequence of an overlap of these spectral components. Such a feature is known for many LBL type blazars \citep[see e.g.][]{Wierzcholska_s5_nus, Wierzcholska_swift}, but has never been reported for \obj\ before.
On the other hand, this flat spectrum can be an effect of an additional spectral component, which can be detected only in the case of the low state of \obj. This can be, for example the Bethe--Heitler emission as proposed by \cite{Petropoulou15}. In such a case, the synchrotron emission from Bethe--Heitler pairs is expected to appear as a “third bump” -- an additional component with a maximum at tens of keV --  in blazar's SED. Its signature can be seen as a lack of a spectral upturn in a broad-band SED.
 
\section{Summary} \label{summary}
\obj\ has been observed by astronomers for many years, but most of the  observations  focused mainly on the optical regime \citep[e.g.][]{Takalo94, Pian95, Sillanpaa96, Pursimo00, Valtonen16}.
This paper focuses on detailed studies of the spectral and temporal variability from optical to X rays of this famous LBL blazar. The long-term observations made with \uvot\ and \xrt\ during the period of 2005--2016 are discussed. 
This work can be summarized as follows:

\begin{itemize}
 \item Significant variability observed in the optical, UV and X-ray ranges has been confirmed in the long-term observations of \obj.  The description of the blazar's behaviour revealed the strongest flux changes observed in the optical/UV regime during the flare observed in 2015 December  and 2016 January. However, in the X-ray regime the simultaneous outburst is not so pronounced. An extraordinary activity of \obj\ over the entire electromagnetic spectrum during the period mentioned has been reported \citep[see e.g.][]{Carrasco_atel, Ciprini_atel, Shappee_atel, Wierzcholska_atel}.
 \item The shortest variability time-scale observed is one day and this is strongly limited by the pointing of the observations.
 \item The variability patterns observed in the optical/UV and X-ray ranges are not correlated.
 \item The long-term optical monitoring of the source has not revealed any clear relation between colour and magnitude. Such a trend has been previously reported but mostly in the case of short-term observations in the high states of the blazar \citep[e.g.][]{Carini1992, Dai2011}.
 \item A harder-when-brighter behaviour is clearly visible in the case of the X-ray data, but not in the case of the optical/UV band.
 \item A statistically significant and strong anti-correlation between the optical/UV and X-ray spectral indices has been found.
 \item In the case of the low state of the blazar observed in 2014--2015 (here interval H), a flat X-ray spectrum has been observed. 
 This feature can be explained as a spectral upturn observed in the X-ray regime or this may be an effect of the influence of an additional spectral component, like the Bethe--Heitler emission, observed in the low state.
\end{itemize}

\begin{table*}  
\caption[]{A characterization of the variability of \obj\ using the X-ray observations. The following columns show: the instrument/filter, the interval, time period in MJD,  the mean flux value given in $10^{-11}\,\textrm{erg}\,\textrm{cm}^{-2}\,\textrm{s}^{-1}$, the reduced chi squares $\chi^2_\textrm{red}$ value of the fit to a constant value and the number of degrees of freedom, the fractional variability amplitude, and two values of the doubling/halving time-scale, as defined in Section~\ref{temporal}, given in days. The intervals are defined in Section~\ref{temporal}, and additionally, we calculate the quantities for the whole data set -- marked as interval 'all'.}
\centering
\begin{tabular}{l|c|c|c|r|c|r|r}
\hline
\hline
Instrument & Interval & Period & Mean value & $\chi^2_\textrm{red}$ & $F_\textrm{var}$ & $\tau_\textrm{d}^{(i)}$ &  $\tau_\textrm{d}^{(ii)}$     \\
\hline
XRT  & A  & 53510.27--54455.02 & $0.25\pm0.02$ &  12.0/28  & $0.311\pm0.022$ & $<$0.01 &  1.61 \\
XRT  & B  & 54763.67--54917.26 & $0.72\pm0.04$ &  10.9/30  & $0.274\pm0.018$ &  1.77 &  1.88 \\
XRT  & C  & 55129.78--55340.81 & $0.41\pm0.03$ &  36.0/24  & $0.409\pm0.019$ &  1.91 &  6.24 \\
XRT  & D  & 55488.02--55647.06 & $0.47\pm0.04$ &  14.9/14  & $0.256\pm0.035$ &  8.94 & 12.64 \\
XRT  & E  & 55878.18--56040.85 & $0.52\pm0.05$ &  38.3/13  & $0.358\pm0.037$ &  6.89 & 12.80 \\
XRT  & F  & 56266.74--56437.64 & $0.36\pm0.02$ &   3.0/12  & $0.119\pm0.038$ & 16.05 & 21.40 \\
XRT  & G  & 56628.38--56775.11 & $0.49\pm0.07$ &  40.0/15  & $0.417\pm0.085$ &  5.59 & 10.70 \\
XRT  & H  & 56992.94--57187.06 & $0.29\pm0.01$ &   7.3/81  & $0.289\pm0.016$ &  0.30 &  0.40 \\
XRT  & I  & 57187.61--57552.93 & $0.45\pm0.02$ &  21.9/45  & $0.357\pm0.016$ &  2.10 &  2.96 \\
XRT  & all& 53510.27--57552.93 & $0.41\pm0.01$ &  31.9/279 & $0.473\pm0.009$ & $<$0.01 &  0.29 \\

\hline
\hline

\end{tabular}
\label{variability_xrt}
\end{table*}

\begin{table*}  
\caption[]{A characterization of the variability of \obj\ using the optical and UV observations. The meaning of the following columns is the same as in Table~\ref{variability_xrt}.}
\centering
\begin{tabular}{l|c|c|r|c|r|r}
\hline
\hline
Instrument & Interval & Mean value & $\chi^2_\textrm{red}$ & $F_\textrm{var}$ & $\tau_\textrm{d}^{(i)}$ &  $\tau_\textrm{d}^{(ii)}$     \\
\hline
UVOT V    & A    & $3.21\pm0.39$ & 259.3/14  & $0.474\pm0.011$ &  8.51 & 17.58 \\
UVOT V    & B    & $2.20\pm0.08$ &  19.6/28  & $0.181\pm0.011$ &  2.80 &  3.78 \\
UVOT V    & C    & $3.19\pm0.14$ &  53.4/20  & $0.194\pm0.009$ &  1.18 & 14.17 \\
UVOT V    & D    & $2.02\pm0.18$ &  78.2/13  & $0.328\pm0.011$ & 18.01 & 26.38 \\
UVOT V    & E    & $2.01\pm0.12$ &  30.5/14  & $0.224\pm0.011$ & 13.64 & 24.49 \\
UVOT V    & F    & $1.61\pm0.20$ &  66.8/10  & $0.406\pm0.015$ & 12.59 & 20.66 \\
UVOT V    & G    & $1.79\pm0.10$ &  19.8/13  & $0.197\pm0.013$ & 13.24 & 21.16 \\
UVOT V    & H    & $2.27\pm0.05$ &  32.5/85  & $0.188\pm0.004$ &  2.59 &  2.84 \\
UVOT V    & I    & $3.42\pm0.26$ &  95.6/22  & $0.368\pm0.008$ &  5.26 &  9.94 \\
UVOT V    & all  & $2.43\pm0.06$ &  76.5/233 & $0.364\pm0.003$ &  1.18 &  2.46 \\
\hline
UVOT B    & A    & $2.90\pm0.36$ & 511.1/13  & $0.462\pm0.008$ & 10.36 & 15.53 \\
UVOT B    & B    & $1.93\pm0.07$ &  40.2/30  & $0.185\pm0.007$ &  1.85 &  2.42 \\
UVOT B    & C    & $3.02\pm0.13$ &  91.1/20  & $0.194\pm0.006$ & 19.36 & 22.36 \\
UVOT B    & D    & $1.85\pm0.14$ & 115.3/15  & $0.298\pm0.008$ & 18.26 & 26.94 \\
UVOT B    & E    & $1.90\pm0.12$ &  58.5/14  & $0.233\pm0.008$ & 11.25 & 23.19 \\
UVOT B    & F    & $1.46\pm0.16$ & 125.5/13  & $0.399\pm0.010$ & 15.11 & 19.13 \\
UVOT B    & G    & $1.68\pm0.09$ &  34.6/13  & $0.194\pm0.009$ & 14.06 & 26.68 \\
UVOT B    & H    & $2.16\pm0.05$ &  70.3/84  & $0.201\pm0.003$ &  2.14 &  2.71 \\
UVOT B    & I    & $3.16\pm0.25$ & 161.2/23  & $0.383\pm0.007$ &  7.13 &  9.09 \\
UVOT B    & all  & $2.25\pm0.05$ & 146.6/240 & $0.367\pm0.002$ &  1.85 &  2.26 \\
\hline
UVOT U    & A    & $2.52\pm0.30$ & 432.3/13  & $0.451\pm0.010$ &  9.54 & 11.75 \\
UVOT U    & B    & $1.59\pm0.05$ &  33.8/30  & $0.187\pm0.007$ &  1.87 &  2.71 \\
UVOT U    & C    & $2.88\pm0.12$ &  73.4/22  & $0.194\pm0.007$ &  5.91 & 19.87 \\
UVOT U    & D    & $1.73\pm0.14$ & 115.9/15  & $0.318\pm0.009$ & 17.01 & 24.19 \\
UVOT U    & E    & $1.82\pm0.10$ &  44.9/16  & $0.225\pm0.009$ &  7.65 & 20.02 \\
UVOT U    & F    & $1.35\pm0.14$ & 106.1/13  & $0.397\pm0.011$ & 14.10 & 18.79 \\
UVOT U    & G    & $1.58\pm0.09$ &  34.7/13  & $0.209\pm0.010$ & 14.52 & 23.79 \\
UVOT U    & H    & $2.12\pm0.04$ &  52.9/86  & $0.193\pm0.003$ &  2.11 &  2.43 \\
UVOT U    & I    & $2.90\pm0.22$ & 113.9/23  & $0.373\pm0.008$ &  5.89 &  8.56 \\
UVOT U    & all  & $2.11\pm0.05$ & 129.7/247 & $0.359\pm0.002$ &  1.87 &  2.17 \\
\hline
UVOT UVW1 & A    & $1.76\pm0.17$ & 338.8/22  & $0.464\pm0.009$ &  3.68 &  5.38 \\
UVOT UVW1 & B    & $1.12\pm0.04$ &  24.8/32  & $0.210\pm0.008$ &  1.87 &  2.47 \\
UVOT UVW1 & C    & $2.30\pm0.10$ &  51.1/23  & $0.199\pm0.009$ &  1.42 &  9.99 \\
UVOT UVW1 & D    & $1.36\pm0.11$ &  97.7/15  & $0.335\pm0.011$ & 13.10 & 22.12 \\
UVOT UVW1 & E    & $1.40\pm0.09$ &  40.0/15  & $0.258\pm0.011$ & 14.74 & 23.61 \\
UVOT UVW1 & F    & $1.05\pm0.11$ &  75.9/13  & $0.406\pm0.013$ & 13.37 & 16.42 \\
UVOT UVW1 & G    & $1.20\pm0.07$ &  31.6/16  & $0.248\pm0.011$ & 12.76 & 21.89 \\
UVOT UVW1 & H    & $1.77\pm0.04$ &  25.9/86  & $0.189\pm0.005$ &  0.80 &  1.66 \\
UVOT UVW1 & I    & $2.35\pm0.17$ & 192.6/44  & $0.487\pm0.007$ &  1.99 &  3.14 \\
UVOT UVW1 & all  & $1.71\pm0.04$ & 129.9/283 & $0.430\pm0.003$ &  0.80 &  1.44 \\
\hline
UVOT UVM2 & A    & $1.59\pm0.18$ & 422.3/17  & $0.487\pm0.008$ &  4.03 &  6.37 \\
UVOT UVM2 & B    & $0.99\pm0.06$ &  33.1/17  & $0.241\pm0.015$ &  1.31 &  2.73 \\
UVOT UVM2 & C    & $2.22\pm0.08$ &  65.8/21  & $0.170\pm0.009$ & 14.72 & 19.11 \\
UVOT UVM2 & D    & $1.33\pm0.13$ & 151.4/13  & $0.368\pm0.010$ & 12.02 & 20.82 \\
UVOT UVM2 & E    & $1.45\pm0.09$ &  43.3/14  & $0.242\pm0.014$ & 18.88 & 25.29 \\
UVOT UVM2 & F    & $1.11\pm0.13$ & 129.8/10  & $0.399\pm0.012$ & 12.64 & 20.78 \\
UVOT UVM2 & G    & $1.19\pm0.08$ &  58.1/17  & $0.289\pm0.009$ &  8.27 & 14.14 \\
UVOT UVM2 & H    & $1.80\pm0.04$ &  34.4/85  & $0.194\pm0.004$ &  0.63 &  1.15 \\
UVOT UVM2 & I    & $2.33\pm0.17$ & 319.0/44  & $0.487\pm0.005$ &  1.51 &  2.17 \\
UVOT UVM2 & all  & $1.73\pm0.05$ & 185.9/253 & $0.428\pm0.003$ &  0.63 &  1.05 \\
\hline
UVOT UVW2 & A    & $1.51\pm0.14$ & 399.5/20  & $0.436\pm0.008$ &  2.90 &  4.83 \\
UVOT UVW2 & B    & $0.91\pm0.05$ &  47.2/19  & $0.227\pm0.010$ &  1.74 &  3.11 \\
UVOT UVW2 & C    & $2.09\pm0.08$ &  72.0/22  & $0.176\pm0.007$ &  1.57 &  9.80 \\
UVOT UVW2 & D    & $1.25\pm0.11$ & 170.1/15  & $0.364\pm0.009$ & 11.42 & 17.95 \\
UVOT UVW2 & E    & $1.38\pm0.08$ &  49.8/14  & $0.233\pm0.009$ & 14.11 & 23.96 \\
UVOT UVW2 & F    & $0.99\pm0.11$ & 102.5/13  & $0.411\pm0.019$ & 13.18 & 16.51 \\
UVOT UVW2 & G    & $1.13\pm0.08$ &  84.1/17  & $0.299\pm0.009$ &  7.00 & 13.88 \\
UVOT UVW2 & H    & $1.70\pm0.04$ &  31.7/87  & $0.194\pm0.004$ &  0.50 &  1.09 \\
UVOT UVW2 & I    & $2.22\pm0.16$ & 261.9/45  & $0.472\pm0.006$ &  1.54 &  2.13 \\
UVOT UVW2 & all  & $1.62\pm0.04$ & 211.7/267 & $0.429\pm0.002$ &  0.50 &  1.05 \\

\hline
\hline

\end{tabular}
\label{variability_uvot}
\end{table*}

\begin{table}
\caption{The parameters of the \po\ spectrum in the X-ray regime for a given interval. The \po\ is defined as: $N (E / 1\,\textrm{keV})^{-\gamma}$, where $N$ is the normalization, $E$ is the energy and $\gamma$ is the index.}
\centering
\begin{tabular}{r|cc}
\hline \hline
Interval & $N$ ($10^{-3}$\,cm$^{-2}$\,s$^{-1}$\,keV$^{-1}$) & $\gamma$ \\
\hline
A & $0.476 \pm 0.008$ & $1.585 \pm 0.020$ \\
B & $1.222 \pm 0.017$ & $1.495 \pm 0.016$ \\
C & $1.369 \pm 0.014$ & $1.811 \pm 0.014$ \\
D & $0.997 \pm 0.020$ & $1.707 \pm 0.026$ \\
E & $1.406 \pm 0.024$ & $1.793 \pm 0.023$ \\
F & $1.088 \pm 0.026$ & $1.763 \pm 0.032$ \\
G & $1.087 \pm 0.021$ & $1.705 \pm 0.026$ \\
H & $1.215 \pm 0.010$ & $2.072 \pm 0.013$ \\
I & $1.220 \pm 0.013$ & $1.755 \pm 0.014$ \\
\hline \hline
\end{tabular}
\label{tab:xrt-spec-intervals}
\end{table}

\section*{Acknowledgements}
A.W. acknowledges support by the Foundation for Polish Science (FNP).
This research was supported in part by PLGrid Infrastructure.
The plots presented in this paper are rendered using Matplotlib \citep{matplotlib}.

\bibliographystyle{mn2e_williams}
\bibliography{references}

\begin{thebibliography}{60}
\expandafter\ifx\csname natexlab\endcsname\relax\def\natexlab#1{#1}\fi

\bibitem[{{Abdo} {et~al}\mbox{.}(2010){Abdo}, {Ackermann}, {Agudo}, {Ajello},
  {Aller}, {Aller}, {Angelakis}, {Arkharov}, {Axelsson}, {Bach}, \&
  et~al.}]{Abdo2010}
{Abdo} A.~A. {et~al.}, 2010, \apj, 716, 30

\bibitem[{{Aharonian} {et~al}\mbox{.}(2007){Aharonian}, {Akhperjanian},
  {Bazer-Bachi}, {Behera}, {Beilicke}, {Benbow}, {Berge}, {Bernl{\"o}hr},
  {Boisson}, {Bolz}, {Borrel}, {Boutelier}, {Braun}, {Brion}, {Brown},
  {B{\"u}hler}, {B{\"u}sching}, {Bulik}, {Carrigan}, {Chadwick}, {Clapson},
  {Chounet}, {Coignet}, {Cornils}, {Costamante}, {Degrange}, {Dickinson},
  {Djannati-Ata{\"i}}, {Domainko}, {Drury}, {Dubus}, {Dyks}, {Egberts},
  {Emmanoulopoulos}, {Espigat}, {Farnier}, {Feinstein}, {Fiasson},
  {F{\"o}rster}, {Fontaine}, {Funk}, {Funk}, {F{\"u}{\ss}ling}, {Gallant},
  {Giebels}, {Glicenstein}, {Gl{\"u}ck}, {Goret}, {Hadjichristidis}, {Hauser},
  {Hauser}, {Heinzelmann}, {Henri}, {Hermann}, {Hinton}, {Hoffmann}, {Hofmann},
  {Holleran}, {Hoppe}, {Horns}, {Jacholkowska}, {de Jager}, {Kendziorra},
  {Kerschhaggl}, {Kh{\'e}lifi}, {Komin}, {Kosack}, {Lamanna}, {Latham}, {Le
  Gallou}, {Lemi{\`e}re}, {Lemoine-Goumard}, {Lenain}, {Lohse}, {Martin},
  {Martineau-Huynh}, {Marcowith}, {Masterson}, {Maurin}, {McComb}, {Moderski},
  {Moulin}, {de Naurois}, {Nedbal}, {Nolan}, {Olive}, {Orford}, {Osborne},
  {Ostrowski}, {Panter}, {Pedaletti}, {Pelletier}, {Petrucci}, {Pita},
  {P{\"u}hlhofer}, {Punch}, {Ranchon}, {Raubenheimer}, {Raue}, {Rayner},
  {Renaud}, {Ripken}, {Rob}, {Rolland}, {Rosier-Lees}, {Rowell}, {Rudak},
  {Ruppel}, {Sahakian}, {Santangelo}, {Saug{\'e}}, {Schlenker}, {Schlickeiser},
  {Schr{\"o}der}, {Schwanke}, {Schwarzburg}, {Schwemmer}, {Shalchi}, {Sol},
  {Spangler}, {Stawarz}, {Steenkamp}, {Stegmann}, {Superina}, {Tam},
  {Tavernet}, {Terrier}, {van Eldik}, {Vasileiadis}, {Venter}, {Vialle},
  {Vincent}, {Vivier}, {V{\"o}lk}, {Volpe}, {Wagner}, {Ward}, \&
  {Zdziarski}}]{2155flare}
{Aharonian} F. {et~al.}, 2007, \apjl, 664, L71

\bibitem[{{Alexander}(1997)}]{Alexander97}
{Alexander} T., 1997, in Astrophysics and Space Science Library, Vol. 218,
  Astronomical Time Series, {Maoz} D., {Sternberg} A., {Leibowitz} E.~M., eds.,
  p. 163

\bibitem[{{Arnaud}(1996)}]{Arnaud96}
{Arnaud} K.~A., 1996, in Astronomical Society of the Pacific Conference Series,
  Vol. 101, Astronomical Data Analysis Software and Systems V, {Jacoby} G.~H.,
  {Barnes} J., eds., p.~17

\bibitem[{{Begelman} {et~al}\mbox{.}(1984){Begelman}, {Blandford}, \&
  {Rees}}]{begelman84}
{Begelman} M.~C., {Blandford} R.~D., {Rees} M.~J., 1984, Reviews of Modern
  Physics, 56, 255

\bibitem[{{B{\"o}ttcher} {et~al}\mbox{.}(2013){B{\"o}ttcher}, {Reimer},
  {Sweeney}, \& {Prakash}}]{Bottcher13}
{B{\"o}ttcher} M., {Reimer} A., {Sweeney} K., {Prakash} A., 2013, \apj, 768, 54

\bibitem[{{Carini} {et~al}\mbox{.}(1992){Carini}, {Miller}, {Noble}, \&
  {Goodrich}}]{Carini1992}
{Carini} M.~T., {Miller} H.~R., {Noble} J.~C., {Goodrich} B.~D., 1992, \aj,
  104, 15

\bibitem[{{Carrasco} {et~al}\mbox{.}(2015){Carrasco}, {Recillas}, {Porras},
  {Chavushyan}, {Leon-Tavares}, \& {Carraminana}}]{Carrasco_atel}
{Carrasco} L., {Recillas} E., {Porras} A., {Chavushyan} V., {Leon-Tavares} J.,
  {Carraminana} A., 2015, The Astronomer's Telegram, 8438

\bibitem[{{Ciprini} {et~al}\mbox{.}(2015){Ciprini}, {Perri}, {Verrecchia}, \&
  {Valtonen}}]{Ciprini_atel}
{Ciprini} S., {Perri} M., {Verrecchia} F., {Valtonen} M., 2015, The
  Astronomer's Telegram, 8401

\bibitem[{{Dai} {et~al}\mbox{.}(2011){Dai}, {Wu}, {Zhu}, {Zhou}, \&
  {Ma}}]{Dai2011}
{Dai} Y., {Wu} J., {Zhu} Z.-H., {Zhou} X., {Ma} J., 2011, \aj, 141, 65

\bibitem[{{Dickel} {et~al}\mbox{.}(1967){Dickel}, {Yang}, {McVittie}, \&
  {Swenson}}]{Dickel67}
{Dickel} J.~R., {Yang} K.~S., {McVittie} G.~C., {Swenson}, Jr. G.~W., 1967,
  \aj, 72, 757

\bibitem[{{Edelson} \& {Krolik}(1988)}]{Edelson1988}
{Edelson} R.~A., {Krolik} J.~H., 1988, \apj, 333, 646

\bibitem[{{Fossati} {et~al}\mbox{.}(1998){Fossati}, {Maraschi}, {Celotti},
  {Comastri}, \& {Ghisellini}}]{fossati98}
{Fossati} G., {Maraschi} L., {Celotti} A., {Comastri} A., {Ghisellini} G.,
  1998, \mnras, 299, 433

\bibitem[{{Gehrels} {et~al}\mbox{.}(2004){Gehrels}, {Chincarini}, {Giommi},
  {Mason}, {Nousek}, {Wells}, {White}, {Barthelmy}, {Burrows}, {Cominsky},
  {Hurley}, {Marshall}, {M{\'e}sz{\'a}ros}, {Roming}, {Angelini}, {Barbier},
  {Belloni}, {Campana}, {Caraveo}, {Chester}, {Citterio}, {Cline}, {Cropper},
  {Cummings}, {Dean}, {Feigelson}, {Fenimore}, {Frail}, {Fruchter}, {Garmire},
  {Gendreau}, {Ghisellini}, {Greiner}, {Hill}, {Hunsberger}, {Krimm},
  {Kulkarni}, {Kumar}, {Lebrun}, {Lloyd-Ronning}, {Markwardt}, {Mattson},
  {Mushotzky}, {Norris}, {Osborne}, {Paczynski}, {Palmer}, {Park}, {Parsons},
  {Paul}, {Rees}, {Reynolds}, {Rhoads}, {Sasseen}, {Schaefer}, {Short},
  {Smale}, {Smith}, {Stella}, {Tagliaferri}, {Takahashi}, {Tashiro},
  {Townsley}, {Tueller}, {Turner}, {Vietri}, {Voges}, {Ward}, {Willingale},
  {Zerbi}, \& {Zhang}}]{Gehrels04}
{Gehrels} N. {et~al.}, 2004, \apj, 611, 1005

\bibitem[{{Ghisellini} {et~al}\mbox{.}(1998){Ghisellini}, {Celotti}, {Fossati},
  {Maraschi}, \& {Comastri}}]{Ghisellini1998}
{Ghisellini} G., {Celotti} A., {Fossati} G., {Maraschi} L., {Comastri} A.,
  1998, \mnras, 301, 451

\bibitem[{{Ghisellini} {et~al}\mbox{.}(2011){Ghisellini}, {Tavecchio},
  {Foschini}, \& {Ghirlanda}}]{Ghisellini2011}
{Ghisellini} G., {Tavecchio} F., {Foschini} L., {Ghirlanda} G., 2011, \mnras,
  414, 2674

\bibitem[{{Giommi} {et~al}\mbox{.}(2006){Giommi}, {Blustin}, {Capalbi},
  {Colafrancesco}, {Cucchiara}, {Fuhrmann}, {Krimm}, {Marchili}, {Massaro},
  {Perri}, {Tagliaferri}, {Tosti}, {Tramacere}, {Burrows}, {Chincarini},
  {Falcone}, {Gehrels}, {Kennea}, \& {Sambruna}}]{Giommi06}
{Giommi} P. {et~al.}, 2006, \aap, 456, 911

\bibitem[{{Gopal-Krishna} {et~al}\mbox{.}(2011){Gopal-Krishna}, {Goyal},
  {Joshi}, {Karthick}, {Sagar}, {Wiita}, {Anupama}, \&
  {Sahu}}]{Gopal-Krishna11}
{Gopal-Krishna}, {Goyal} A., {Joshi} S., {Karthick} C., {Sagar} R., {Wiita}
  P.~J., {Anupama} G.~C., {Sahu} D.~K., 2011, \mnras, 416, 101

\bibitem[{{Gupta} {et~al}\mbox{.}(2012){Gupta}, {Pandey}, {Singh}, {Rani},
  {Pan}, {Fan}, \& {Gupta}}]{Gupta12}
{Gupta} S.~P., {Pandey} U.~S., {Singh} K., {Rani} B., {Pan} J., {Fan} J.~H.,
  {Gupta} A.~C., 2012, \na, 17, 8

\bibitem[{{H.E.S.S.~Collaboration}(2014)}]{Abramowski2014}
{H.E.S.S.~Collaboration}, 2014, \aap, 571, A39

\bibitem[{Hunter(2007)}]{matplotlib}
Hunter J.~D., 2007, Computing In Science \& Engineering, 9, 90

\bibitem[{{Ikejiri} {et~al}\mbox{.}(2011){Ikejiri}, {Uemura}, {Sasada}, {Ito},
  {Yamanaka}, {Sakimoto}, {Arai}, {Fukazawa}, {Ohsugi}, {Kawabata}, {Yoshida},
  {Sato}, \& {Kino}}]{Ikejiri_2011}
{Ikejiri} Y. {et~al.}, 2011, \pasj, 63, 639

\bibitem[{Jones {et~al}\mbox{.}(2001)Jones, Oliphant, Peterson,
  {et~al.}}]{scipy}
Jones E., Oliphant T., Peterson P., {et~al.}, 2001, {SciPy}: Open source
  scientific tools for {Python}. [Online; accessed 2016-08-23]

\bibitem[{{Kalberla} {et~al}\mbox{.}(2005){Kalberla}, {Burton}, {Hartmann},
  {Arnal}, {Bajaja}, {Morras}, \& {P{\"o}ppel}}]{Kalberla05}
{Kalberla} P.~M.~W., {Burton} W.~B., {Hartmann} D., {Arnal} E.~M., {Bajaja} E.,
  {Morras} R., {P{\"o}ppel} W.~G.~L., 2005, \aap, 440, 775

\bibitem[{{Kirk} {et~al}\mbox{.}(1998){Kirk}, {Rieger}, \&
  {Mastichiadis}}]{Kirk98}
{Kirk} J.~G., {Rieger} F.~M., {Mastichiadis} A., 1998, \aap, 333, 452

\bibitem[{{Maraschi} {et~al}\mbox{.}(1992){Maraschi}, {Ghisellini}, \&
  {Celotti}}]{Maraschi92}
{Maraschi} L., {Ghisellini} G., {Celotti} A., 1992, \apjl, 397, L5

\bibitem[{{Meyer} {et~al}\mbox{.}(2011){Meyer}, {Fossati}, {Georganopoulos}, \&
  {Lister}}]{Meyer11}
{Meyer} E.~T., {Fossati} G., {Georganopoulos} M., {Lister} M.~L., 2011, \apj,
  740, 98

\bibitem[{{M{\"u}cke} {et~al}\mbox{.}(2003){M{\"u}cke}, {Protheroe}, {Engel},
  {Rachen}, \& {Stanev}}]{Mucke13}
{M{\"u}cke} A., {Protheroe} R.~J., {Engel} R., {Rachen} J.~P., {Stanev} T.,
  2003, Astroparticle Physics, 18, 593

\bibitem[{{Padovani} \& {Giommi}(1995)}]{padovani95}
{Padovani} P., {Giommi} P., 1995, \apj, 444, 567

\bibitem[{{Petropoulou} \& {Mastichiadis}(2015)}]{Petropoulou15}
{Petropoulou} M., {Mastichiadis} A., 2015, \mnras, 447, 36

\bibitem[{{Pian} {et~al}\mbox{.}(1995){Pian}, {Edelson}, {Wagner}, {Bregman},
  {George}, {Treves}, {Wamsteker}, {Bock}, {Carini}, {Courvoisier}, {Donahue},
  {Efimov}, {Filippenko}, {Fink}, {Heidt}, {Lawrence}, {Maraschi}, {Miller},
  {Pike}, {Quirrenbach}, {Shakhovskoy}, {Sillanp{\aa}{\aa}}, {Sitko}, {Smith},
  {Takalo}, {Ter{\aa}sranta}, {Valtaoja}, {Ward}, \& {Warwick}}]{Pian95}
{Pian} E. {et~al.}, 1995, Advances in Space Research, 16, 57

\bibitem[{{Pian} {et~al}\mbox{.}(1998){Pian}, {Vacanti}, {Tagliaferri},
  {Ghisellini}, {Maraschi}, {Treves}, {Urry}, {Fiore}, {Giommi}, {Palazzi},
  {Chiappetti}, \& {Sambruna}}]{Pian88}
{Pian} E. {et~al.}, 1998, \apjl, 492, L17

\bibitem[{{Poole} {et~al}\mbox{.}(2008){Poole}, {Breeveld}, {Page}, {Landsman},
  {Holland}, {Roming}, {Kuin}, {Brown}, {Gronwall}, {Hunsberger}, {Koch},
  {Mason}, {Schady}, {vanden Berk}, {Blustin}, {Boyd}, {Broos}, {Carter},
  {Chester}, {Cucchiara}, {Hancock}, {Huckle}, {Immler}, {Ivanushkina},
  {Kennedy}, {Marshall}, {Morgan}, {Pandey}, {de Pasquale}, {Smith}, \&
  {Still}}]{Poole08}
{Poole} T.~S. {et~al.}, 2008, \mnras, 383, 627

\bibitem[{{Poutanen} {et~al}\mbox{.}(2008){Poutanen}, {Zdziarski}, \&
  {Ibragimov}}]{Poutanen08}
{Poutanen} J., {Zdziarski} A.~A., {Ibragimov} A., 2008, \mnras, 389, 1427

\bibitem[{{Pursimo} {et~al}\mbox{.}(2000){Pursimo}, {Takalo},
  {Sillanp{\"a}{\"a}}, {Kidger}, {Lehto}, {Heidt}, {Charles}, {Aller}, {Aller},
  {Beckmann}, {Ben{\'{\i}}tez}, {Bock}, {Boltwood}, {Borgeest}, {de Diego}, {De
  Francesco}, {Dietrich}, {Dultzin-Hacyan}, {Efimov}, {Fiorucci}, {Ghisellini},
  {Gonz{\'a}lez-P{\'e}rez}, {Hanski}, {Hein{\"a}m{\"a}ki}, {Honeycutt},
  {Hughes}, {Karlamaa}, {Katajainen}, {Knee}, {Kurtanidze}, {K{\"u}mmel},
  {K{\"u}hl}, {Lainela}, {Lanteri}, {Linde}, {L{\"a}hteenm{\"a}ki}, {Maesano},
  {Mahoney}, {Marchenko}, {Marscher}, {Massaro}, {Montagni}, {Nesci},
  {Nikolashvili}, {Nilsson}, {Nurmi}, {Pietil{\"a}}, {Poyner}, {Raiteri},
  {Rekola}, {Richter}, {Riehokainen}, {Robertson}, {Rodr{\'{\i}}guez-Espinoza},
  {Sadun}, {Shakhovskoy}, {Schramm}, {Schramm}, {Sobrito}, {Teerikorpi},
  {Ter{\"a}sranta}, {Tornikoski}, {Tosti}, {Turner}, {Valtaoja}, {Valtonen},
  {Villata}, {Wagner}, {Webb}, {Weneit}, \& {Wiren}}]{Pursimo00}
{Pursimo} T. {et~al.}, 2000, \aaps, 146, 141

\bibitem[{{Sagar} {et~al}\mbox{.}(2004){Sagar}, {Stalin}, {Gopal-Krishna}, \&
  {Wiita}}]{Sagar04}
{Sagar} R., {Stalin} C.~S., {Gopal-Krishna}, {Wiita} P.~J., 2004, \mnras, 348,
  176

\bibitem[{{Saito} {et~al}\mbox{.}(2013){Saito}, {Stawarz}, {Tanaka},
  {Takahashi}, {Madejski}, \& {D'Ammando}}]{saito}
{Saito} S., {Stawarz} {\L}., {Tanaka} Y.~T., {Takahashi} T., {Madejski} G.,
  {D'Ammando} F., 2013, \apjl, 766, L11

\bibitem[{{Schlafly} \& {Finkbeiner}(2011)}]{Schlafly11}
{Schlafly} E.~F., {Finkbeiner} D.~P., 2011, \apj, 737, 103

\bibitem[{{Shappee} {et~al}\mbox{.}(2015){Shappee}, {Stanek}, {Holoien},
  {Brown}, {Kochanek}, {Godoy-Rivera}, {Basu}, {Prieto}, {Bersier}, {Dong},
  {Chen}, \& {Brimacombe}}]{Shappee_atel}
{Shappee} B.~J. {et~al.}, 2015, The Astronomer's Telegram, 8372

\bibitem[{{Sikora} {et~al}\mbox{.}(1994){Sikora}, {Begelman}, \&
  {Rees}}]{Sikora94}
{Sikora} M., {Begelman} M.~C., {Rees} M.~J., 1994, \apj, 421, 153

\bibitem[{{Sillanpaa} {et~al}\mbox{.}(1988){Sillanpaa}, {Haarala}, {Valtonen},
  {Sundelius}, \& {Byrd}}]{Sillanpaa88}
{Sillanpaa} A., {Haarala} S., {Valtonen} M.~J., {Sundelius} B., {Byrd} G.~G.,
  1988, \apj, 325, 628

\bibitem[{{Sillanpaa} {et~al}\mbox{.}(1996){Sillanpaa}, {Takalo}, {Pursimo},
  {Nilsson}, {Heinamaki}, {Katajainen}, {Pietila}, {Hanski}, {Rekola},
  {Kidger}, {Boltwood}, {Turner}, {Robertson}, {Honeycut}, {Efimov},
  {Shakhovskoy}, {Fiorucci}, {Tosti}, {Ghisellini}, {Raiteri}, {Villata}, {de
  Francesco}, {Lanteri}, {Chiaberge}, {Peila}, \& {Heidt}}]{Sillanpaa96}
{Sillanpaa} A. {et~al.}, 1996, \aap, 315, L13

\bibitem[{{Takalo} {et~al}\mbox{.}(1994){Takalo}, {Sillanpaeae}, \&
  {Nilsson}}]{Takalo94}
{Takalo} L.~O., {Sillanpaeae} A., {Nilsson} K., 1994, \aaps, 107

\bibitem[{{Urry} \& {Padovani}(1995)}]{Urry95}
{Urry} C.~M., {Padovani} P., 1995, \pasp, 107, 803

\bibitem[{{Valtonen} {et~al}\mbox{.}(2006){Valtonen}, {Lehto},
  {Sillanp{\"a}{\"a}}, {Nilsson}, {Mikkola}, {Hudec}, {Basta},
  {Ter{\"a}sranta}, {Haque}, \& {Rampadarath}}]{Valtonen06}
{Valtonen} M.~J. {et~al.}, 2006, \apj, 646, 36

\bibitem[{{Valtonen} \& {Wiik}(2012)}]{Valtonen12}
{Valtonen} M.~J., {Wiik} K., 2012, \mnras, 421, 1861

\bibitem[{{Valtonen} {et~al}\mbox{.}(2016){Valtonen}, {Zola}, {Ciprini},
  {Gopakumar}, {Matsumoto}, {Sadakane}, {Kidger}, {Gazeas}, {Nilsson},
  {Berdyugin}, {Piirola}, {Jermak}, {Baliyan}, {Alicavus}, {Boyd}, {Campas
  Torrent}, {Campos}, {Carrillo G{\'o}mez}, {Caton}, {Chavushyan}, {Dalessio},
  {Debski}, {Dimitrov}, {Drozdz}, {Er}, {Erdem}, {Escartin P{\'e}rez}, {Fallah
  Ramazani}, {Filippenko}, {Ganesh}, {Garcia}, {G{\'o}mez Pinilla},
  {Gopinathan}, {Haislip}, {Hudec}, {Hurst}, {Ivarsen}, {Jelinek}, {Joshi},
  {Kagitani}, {Kaur}, {Keel}, {LaCluyze}, {Lee}, {Lindfors}, {Lozano de Haro},
  {Moore}, {Mugrauer}, {Naves Nogues}, {Neely}, {Nelson}, {Ogloza}, {Okano},
  {Pandey}, {Perri}, {Pihajoki}, {Poyner}, {Provencal}, {Pursimo}, {Raj},
  {Reichart}, {Reinthal}, {Sadegi}, {Sakanoi}, {Salto Gonz{\'a}lez}, {Sameer},
  {Schweyer}, {Siwak}, {Sold{\'a}n Alfaro}, {Sonbas}, {Steele}, {Stocke},
  {Strobl}, {Takalo}, {Tomov}, {Tremosa Espasa}, {Valdes}, {Valero P{\'e}rez},
  {Verrecchia}, {Webb}, {Yoneda}, {Zejmo}, {Zheng}, {Telting}, {Saario},
  {Reynolds}, {Kvammen}, {Gafton}, {Karjalainen}, {Harmanen}, \&
  {Blay}}]{Valtonen16}
{Valtonen} M.~J. {et~al.}, 2016, \apjl, 819, L37

\bibitem[{{Vaughan} {et~al}\mbox{.}(2003{\natexlab{a}}){Vaughan}, {Edelson},
  {Warwick}, \& {Uttley}}]{Vaughan2003}
{Vaughan} S., {Edelson} R., {Warwick} R.~S., {Uttley} P., 2003{\natexlab{a}},
  \mnras, 345, 1271

\bibitem[{{Vaughan} {et~al}\mbox{.}(2003{\natexlab{b}}){Vaughan}, {Edelson},
  {Warwick}, \& {Uttley}}]{Vaughan03}
{Vaughan} S., {Edelson} R., {Warwick} R.~S., {Uttley} P., 2003{\natexlab{b}},
  \mnras, 345, 1271

\bibitem[{{Wagner}(2009)}]{Wagner2009}
{Wagner} S., 2009, in Astrophysics with All-Sky X-Ray Observations, {Kawai} N.,
  {Mihara} T., {Kohama} M., {Suzuki} M., eds., p. 186

\bibitem[{{Wagner} \& {Witzel}(1995)}]{wagner}
{Wagner} S.~J., {Witzel} A., 1995, \araa, 33, 163

\bibitem[{{Wierzcholska}(2015)}]{Wierzcholska_48}
{Wierzcholska} A., 2015, \aap, 580, A104

\bibitem[{{Wierzcholska} {et~al}\mbox{.}(2015){Wierzcholska}, {Ostrowski},
  {Stawarz}, {Wagner}, \& {Hauser}}]{Wierzcholska_atom}
{Wierzcholska} A., {Ostrowski} M., {Stawarz} {\L}., {Wagner} S., {Hauser} M.,
  2015, \aap, 573, A69

\bibitem[{{Wierzcholska} \& {Siejkowski}(2015{\natexlab{a}})}]{Wierzcholska_s5}
{Wierzcholska} A., {Siejkowski} H., 2015{\natexlab{a}}, \mnras, 452, L11

\bibitem[{{Wierzcholska} \&
  {Siejkowski}(2015{\natexlab{b}})}]{Wierzcholska_atel}
{Wierzcholska} A., {Siejkowski} H., 2015{\natexlab{b}}, The Astronomer's
  Telegram, 8395

\bibitem[{{Wierzcholska} \& {Siejkowski}(2016)}]{Wierzcholska_s5_nus}
{Wierzcholska} A., {Siejkowski} H., 2016, \mnras, 458, 2350

\bibitem[{{Wierzcholska} \& {Wagner}(2016)}]{Wierzcholska_swift}
{Wierzcholska} A., {Wagner} S.~J., 2016, \mnras, 458, 56

\bibitem[{{Wu} {et~al}\mbox{.}(2006){Wu}, {Zhou}, {Wu}, {Liu}, {Peng}, {Ma},
  {Wu}, {Jiang}, \& {Chen}}]{Wu06}
{Wu} J. {et~al.}, 2006, \aj, 132, 1256

\bibitem[{{Zhang} {et~al}\mbox{.}(1999){Zhang}, {Celotti}, {Treves},
  {Chiappetti}, {Ghisellini}, {Maraschi}, {Pian}, {Tagliaferri}, {Tavecchio},
  \& {Urry}}]{Zhang99}
{Zhang} Y.~H. {et~al.}, 1999, \apj, 527, 719

\bibitem[{{Zhang} {et~al}\mbox{.}(2005){Zhang}, {Treves}, {Celotti}, {Qin}, \&
  {Bai}}]{Zhang2005}
{Zhang} Y.~H., {Treves} A., {Celotti} A., {Qin} Y.~P., {Bai} J.~M., 2005, \apj,
  629, 686

\end{thebibliography}

\label{lastpage}

\end{document}